\pdfoutput=1

\RequirePackage{fix-cm}
\documentclass[twocolumn,epjc3]{svjour3}  

\usepackage[style=phys,eprint=true,url=true]{biblatex}
\RequirePackage{graphicx}
\RequirePackage{mathptmx}
\RequirePackage{float}
\RequirePackage{enumitem}
\RequirePackage[separate-uncertainty=true]{siunitx}
\RequirePackage{xspace}
\RequirePackage{amsmath}
\RequirePackage{subcaption}
\RequirePackage{doi}
\usepackage[pagewise]{lineno}

\newcommand{\ptr}{\texttt{ptr}\xspace}
\newcommand{\absz}{$z_{\text{abs}}$\xspace}
\newcommand{\lst}{$R_T$\xspace}
\journalname{Eur. Phys. J. C}
\begin{document}

\title{Primary track recovery in high-definition gas time projection chambers}

\author{P. M. Lewis\thanksref{e1,ubonn}
  \and
  M. T. Hedges\thanksref{purdue}
  \and
  I. Jaegle\thanksref{e2,uh}
  \and
  J. Schueler\thanksref{uh}
  \and
  T. N. Thorpe\thanksref{e4,uh}
  \and
  S. E. Vahsen\thanksref{uh}
}

\thankstext{e1}{e-mail: lewis@physik.uni-bonn.de}
\thankstext{e2}{Now at Thomas Jefferson National Accelerator Facility, Newport News, VA, 23606, USA}
\thankstext{e4}{Now at University of California Los Angeles, Los Angeles, CA 90095-1547, USA}

\institute{University of Bonn, Institute of Physics, Nu\ss{}allee 12, 53115 Bonn, Germany\label{ubonn}
           \and
           Purdue University, Department of Physics and Astronomy, 525 Northwestern Ave, West Lafayette, IN 47907, USA \label{purdue}
           \and
           University of Hawaii, Department of Physics and Astronomy, 2505 Correa Road, Honolulu, HI 96822, USA \label{uh}
}

\date{Received: date / Accepted: date}

\maketitle

\begin{abstract}
We develop and validate a new algorithm called \textit{primary track recovery} (\ptr) that effectively deconvolves known physics and detector effects from nuclear recoil tracks in gas time projection chambers (TPCs) with high-resolution readout. This gives access to the primary track charge, length, and vector direction (helping to resolve the ``head-tail'' ambiguity). Additionally, \ptr provides a measurement of the transverse and longitudinal diffusion widths, which can be used to determine the absolute position of tracks in the drift direction for detector fiducialization. Using simulated helium recoils in an atmospheric pressure TPC with a 70:30 mixture of He:CO$_2$ we compare the performance of \ptr to traditional methods for all key track variables. We find that the algorithm reduces reconstruction errors, including those caused by charge integration, for tracks with mean length-to-width ratios 1.4 and above, corresponding to recoil energies of $20$~keV and above in the studied TPCs. We show that \ptr improves on existing methods for head-tail disambiguation, particularly for highly inclined tracks, and improves the determination of the absolute position of recoils on the drift axis via transverse diffusion. We find that \ptr can partially recover charge structure integrated out by the detector in the $z$ direction, but that its determination of energy and length have worse resolution compared to existing methods. We use experimental data to qualitatively verify these findings and discuss implications for future directional detectors at the low-energy frontier. 

\keywords{TPC \and pixels \and head-tail \and absolute position \and absolute z \and neutron \and directional \and dark matter \and MPGD \and nuclear recoils}
\end{abstract}

\section{Introduction}
Gas Time Projection Chambers with high-resolution amplification and readout (``HD TPCs'') are ideally suited to the directional detection of low-energy nuclear recoil tracks~\cite{vahsen_3d}. This capability has been used in statistical neutron tracking and may permit dark matter searches below the neutrino floor~\cite{spergel,grothaus,ohare,mayet,cygnus,ddm_review}. In these directional dark matter searches, extending the low-energy limit of directional sensitivity could unlock a region of low-mass dark matter that is inaccessible to detectors without directional sensitivity.

The technological key to low-energy recoil tracking is amplification and readout with high resolution in both space and time~\cite{battat}. However, many such detectors integrate charge in time at the hit level, leading to a loss of charge distribution information along the drift axis. We refer to this as \textit{2.5D} reconstruction. Our primary aim is to develop a track processing algorithm that restores the lost charge distribution information, effectively converting 2.5D reconstructed tracks into 3D recovered tracks. 

Independent of integration effects, the low-energy limit for directionality ultimately will be dictated by thermal diffusion, particularly in large-volume TPCs. Therefore, it is essential that any advanced track processing algorithm is robust against diffusion. 

Here we argue for, develop, and demonstrate a new approach to processing recoil tracks in HD TPCs, called \textit{primary track recovery} (\ptr). Conceptually, \ptr can be thought of as a method to effectively deconvolve a detected 2.5D track with an adaptive kernel that encodes all drift, amplification, and digitization dynamics. In this way, the full 3D primary charge distribution can be largely recovered, leading to improved determination of collective properties of the primary track, including the angles $\theta$ and $\phi$ without ``head-tail'' ambiguity, absolute drift distance \absz, length $L$, and charge $q$. Although our validation uses electron-drift atmospheric-pressure gas TPCs~\cite{beast_tpcs} with gas electron multipliers (GEMs)~\cite{gem} and pixel readout, we develop the technique in a general way that could be easily adapted for other detectors, for example HD TPCs operating at low pressure or with negative ion drift (NID)~\cite{thorpe}. For this reason, we do not focus on absolute performance, but rather improvements relative to existing techniques for a given detector.

In Sec.~\ref{sec:intro_primary_track_properties} we give an overview of the primary track properties, including their use and the current status of their measurement. In Sec.~\ref{sec:intro_simulation} we describe the experimental and simulated TPCs used in subsequent studies. In Sec.~\ref{sec:model} we develop a continuous model of track development and in Sec.~\ref{sec:technique} we use this model to build the \ptr algorithm. We then use simulated and experimental data to test the performance of \ptr in Sec.~\ref{sec:performance} and Sec.~\ref{sec:demonstrations}, respectively. Finally, in Sec.~\ref{sec:conclusions} we summarize our results and discuss implications. 

\section{\label{sec:intro_primary_track_properties}Primary track properties}
A primary track fundamentally consists of the 3D coordinates of every electron generated by a recoiling nucleus. However, it is the bulk properties of this distribution that are generally most connected to the underlying physics. Below we summarize each of these primary track properties, why they are useful, the current techniques for and challenges with measuring them, and why \ptr may allow for an improvement over existing techniques.

In order to facilitate easy comparison between detectors, we evaluate track reconstruction performance in terms of the \textit{transverse aspect ratio} of a primary track, approximated as \lst$=(L+4\sigma_T)/4\sigma_T$, where $L$ is the length of the initial ionization trail and $\sigma_T$ is the diffusion width of the detected track in the transverse plane. This approximation assumes that the edges of the track are $2\sigma_T$ away from the track center. For a given recoil energy, $L$ and $\sigma_T$ will vary independently based on detector design and gas properties, but we expect similar performance between different HD detectors for a fixed value of \lst. In particular, we use \lst$=2$ as a benchmark; at this working point the detected tracks are roughly twice as long as they are wide, which could be the case for a medium-energy track in a high-pressure, low-gain, electron-drift TPC, or a very low-energy track in a low-pressure, high-gain, NID TPC. 

\subsection{Energy, charge, and length}
The \textit{charge} $q$ of a primary track is the number of electrons comprising it. This is related to the \textit{ionization energy} $E$ via the work function of the gas. The ionization energy is related to the recoil energy via the nuclear quenching factor, which approaches unity for higher recoil energies. Therefore, our recovered energy, measured in keV, differs from the recoil energy, especially at low energies. 

The \textit{length} $L$ is the distance between the first and last ionizations in the primary track, which is a proxy for the total distance the recoil travels. Although this distance in principle includes many deflections due to multiple elastic scattering, in practice it is generally sufficient to consider the length of the primary track projected on its major axis.

Together, the charge/energy and length determine the stopping power $\text{d}E/\text{d}x$, which can be used to discriminate between different recoiling species~\cite{hedges}, and also between nuclear and electronic recoils. For very low-energy applications such as directional dark matter searches, electron recoils constitute the dominant background, so a premium is placed on the accuracy and resolution of $q$ and $L$. The energy can be independently important in the case that a recoil spectrum is desired, for example in the case of reconstructing the energy of incoming neutrons or neutrinos from a known direction~\cite{ddm_review}. 

The essential obstacle to an accurate length measurement is diffusion, which becomes more important for small values of \lst; at \lst$=2$ the error on the length measurement is roughly $100\%$. Unfortunately, low-length and low-energy tracks typically constitute both the most common and the most important signal. 

Errors in charge measurement can come from multiple sources: fluctuations in or mismeasurements of the gain, charge lost through recombination or electron attachment, and charge lost under threshold or due to saturation of the detector. Of these effects, we focus on those caused by digitization, as the others can be controlled with calibration. Because \ptr is designed to deconvolve all of the effects that lead to length and charge errors, we expect it to improve the accuracy of these parameters.

\subsection{The angles $\theta$ and $\phi$}\label{sec:intro_angles}
Directional detection necessarily includes determination of the initial recoil direction, which we obtain from the spatial distribution of the primary charges. For this, we define the \textit{track vector} as a unit vector that points along the major axis of the track, in the direction the recoil traveled (toward the ``head'' of the track). For sufficiently long tracks, the polar angle of the track vector with respect to the drift axis $\theta$ and the azimuthal angle $\phi$ can be extracted by a simple linear fit to the detected hit distribution. However, this fit cannot determine the \textit{vector direction} of a track; it can't determine whether the track vector or its 3D reflection ($\vec{v}$ vs. $-\vec{v}$) points to the head of the track. 

In some applications, for example directional dark matter observatories~\cite{cygnus}, it is essential to be able to distinguish the head of a track from its ``tail''~\cite{driftIIc}. Previous studies~\cite{cygnus} have shown that extending directionality to keV-scale recoil energies is highly desirable for improved sensitivity to low-mass dark matter models and solar neutrinos. Resolution of \textit{head-tail} ambiguity on a track-by-track basis at low recoil energies is therefore highly desirable. A detector with such a capability is considered to have vector directionality. In the past, vector directionality has been achieved by exploiting the charge imbalance between the head and the tail of tracks~\cite{driftIId}\cite{dmtpc_ht}. Recently, NEWAGE-0.3b'~\cite{newage} reported the first use of 3D vector directionality in a dark matter search, with Fluorine recoils down to ionization energies of 50~keV in CF$_4$ gas at 76~torr. Subsequently, we submitted the first demonstration of 3D vector tracking of He recoils, down to 100~keV in He:CO$_2$ at one atmosphere~\cite{hedges}, which is at \lst$\approx2.5$. However, we will show here that integration effects degrade the performance of this technique for inclined tracks, and significant improvements are possible through understanding, modeling, and deconvolving these effects. We expect deconvolution of diffusion to permit high-efficiency head-tail determination at \lst$=2$ or below.  

Additionally, a directional detector needs to be able to identify the direction of a source of particles with a resolution comparable to or better than the typical elastic recoil angle, apart from head-tail swaps. For \lst$=2.5$ and above, we have demonstrated angular resolutions better than $20^{\circ}$, which is sufficient for point source identification~\cite{hedges}. However, at sufficiently low \lst, the combined effects of diffusion and transverse straggling render the angular resolution inadequate for tracking. This is, fundamentally, a detector challenge and we do not expect to be able to improve significantly algorithmically. However, we will show that \ptr provides a potential means to improve angular reconstruction when it is limited more by transverse straggling than by diffusion.

\subsection{The position $(x,y,z)$}
In low-background applications, charged particles emitted by radioactive decay in detector material outside the active volume are a major source of background. The precise location of the primary tracks on the readout plane $(x,y)$ can then be used to \textit{fiducialize} the volume by removing any tracks that cross the edge of the sensitive volume. However, this cannot be done in $z$ because the absolute position of the track in $z$, \textit{absolute z} (\absz), is not intrinsically measured. 

We have previously shown~\cite{absz} that it is possible to estimate \absz for high-energy alpha tracks by measuring the transverse width of the tracks, which grows like $\sqrt{z_{\text{abs}}}$. Using 8~mm alpha track segments, we found that \absz can be determined on a track-by-track basis with a resolution better than 1~cm for drift lengths less than 10~cm. We argued that this technique could be used to fiducialize the TPC volume by rejecting recoil candidates with widths consistent with emission from the cathode. Subsequently, this was demonstrated for the first time for nuclear recoils with electron drift in a small TPC~\cite{minitpc} utilizing the same charge readout scheme as ours. In the process of deconvolving the diffusion, \ptr can provide a precision measurement of it, thereby permitting \absz determination to high precision for low-energy nuclear recoils.

\subsection{\label{sec:intro_other_track_properties}Other track properties}
Although not primary track properties, the widths $\sigma_T$ and $\sigma_L$ are determined by \ptr as part of adapting the deconvolution kernel to the track. The transverse width $\sigma_T$ is used to determine \absz. In principle, the same could be done using $\sigma_L$; however, we will see that this is considerably more difficult. Although further applications are not clear, \ptr will provide the first means to measure both the transverse and longitudinal diffusion of nuclear recoils on a track-by-track basis. 

\section{\label{sec:intro_simulation}Simulated and physical TPCs}
In the validation of \ptr we use simulated and experimental datasets. In both cases, for convenience, we use the BEAST TPCs \cite{beast_tpcs} operating in low-gain mode optimized for fast neutron detection. We describe these TPCs and their operational parameters below. We use these TPCs for demonstration purposes only; in fast-neutron mode, low-energy performance is not ideal, and furthermore the performance of \ptr as a function of recoil energy is not externally meaningful. We focus instead on improvements relative to existing reconstruction methods.

We use the same simulation methodology that we have used in previous work. We have found general agreement between data and simulation for key observables such as point resolution, diffusion, angular resolution, length, and energy~\cite{vahsen_3d,hedges}. However, \ptr relies on modeling features of track development that are not exploited by previous methods. Consequently, new sources of disagreement may arise. We present the first tests of agreement in Sec.~\ref{sec:demonstrations}.

\subsection{\label{sec:intro_tpcs}The BEAST TPCs}
The BEAST TPCs consist of a $2.0 \times 1.68 \times \SI{10.87}{\centi\meter\cubed}$ active volume of a 70:30 mixture of atmospheric pressure He:CO$_2$. A series of aluminum rings provide a uniform electric drift field of $\SI{530}{\volt\per\centi\meter}$, resulting in a drift velocity of $\SI{250}{\micro\meter}/\SI{25}{\nano\second}$ for the simulated TPC. The drift velocity in the physical TPC used in the validation is $\SI{216}{\micro\meter\per\num{25~}\nano\second}$. A pair of GEMs mounted below the field cage amplifies charge by roughly a factor of 40 each, resulting in a gain of roughly 1600 when operating at low gains suitable for the detection of nuclear recoils. We use a FE-I4b~\cite{fei4b} pixel chip to digitize the amplified charge with high resolution ($250\times\SI{50}{\micro\meter\squared}$ pixels digitized at 40~MHz, with 4 bits of time-over threshold (TOT) information per pixel). The effective gain for the simulated (physical) TPC is $G=1416(883)$ with a threshold of 2000 electrons and a saturation level of $43,680(47,304)$ electrons.

\subsection{\label{sec:sim_srim}Primary ionization}
In order to simulate primary ionization, we use the event generator SRIM \cite{srim} with default quenching factor values. We use the same method as in Ref.~\cite{cygnus} and Ref.~\cite{cosmin} to simulate both primary and secondary ionizations, using values for the work function ($w=\SI{35.075}{\eV}$) and Fano factor ($0.19$) estimated using GARFIELD++ \cite{garfield}. The result of this simulation is a 3D distribution of electrons, which we pass on to the drift and amplification simulation. 

\subsection{\label{sec:sim_drift_and_amplification}Drift and amplification}
To simulate the effects of diffusion and resolution smearing during drift, we use a model based on Gaussian effective resolutions, as described in Ref.~\cite{vahsen_3d}. We randomize each electron's coordinates in $x-y$ using a Gaussian distribution with width $\sigma_T=\sqrt{B_Tz}$ and in $z$ with width $\sigma_L=\sqrt{B_Lz}$, where $z$ is the initial position of the primary charge above the amplification plane. We estimate values for the \textit{diffusion parameters} $B_T$ and $B_L$, as well as the drift velocity $v_D$, using Magboltz 2~\cite{magboltz1,magboltz2}. The diffusion parameters are related to the diffusion constants $D$ and drift velocity via $B=\sqrt{2D/v_D}$.

\begin{figure}
  \includegraphics[width=0.9\columnwidth]{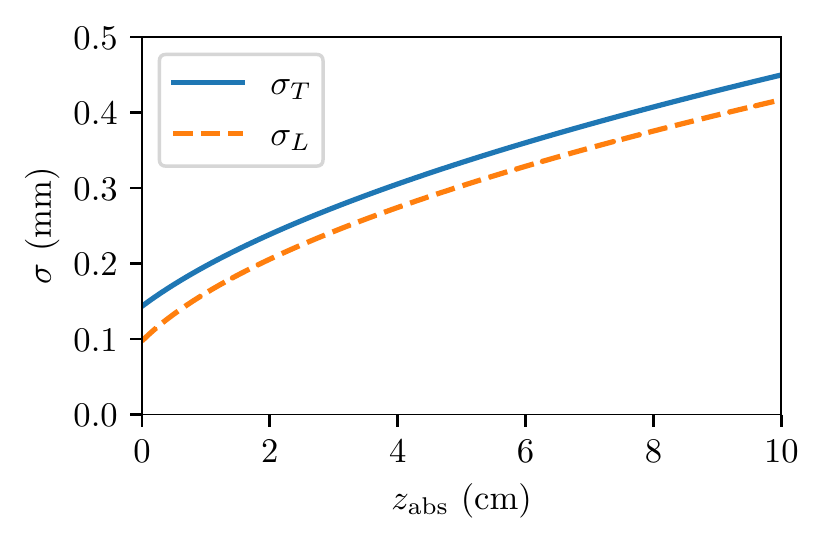}
  \caption{The dependence of the transverse width $\sigma_T$ (solid blue line) and the longitudinal width $\sigma_L$ (dashed orange line) on the drift distance \absz for the simulated TPC. The offset at \absz$=0$ is due to the amplification resolution term $A_T$. The measured widths will include an additional contribution from the digitization resolution of the readout.}
\label{fig:sigma_vs_absz}
\end{figure}

We simulate avalanche gain at the single electron level using an exponential gain distribution. Individual electrons after gain are again smeared in space individually using a combined amplification resolution. In the transverse direction, this resolution includes contributions from quantization into two GEM holes, and diffusion in the gaps between the GEMs and between the bottom GEM and the pixel chip. In the longitudinal direction, this resolution includes diffusion in the gaps and additional diffusion arising during the avalanche process. We estimate the diffusion and avalanche contributions with Magboltz. These effects lead to ``fixed'' transverse and longitudinal widths independent of the drift distance, $A_T$ and $A_L$, so the combined widths can be written as:
\begin{align}
  \sigma_T &= \sqrt{A^2_T+B^2_Tz}\label{eq:sigT},\\
  \sigma_L &= \sqrt{A^2_L+B^2_Lz}\label{eq:sigL}.
\end{align}
The resulting dependence of the widths on the drift distance is shown in Fig.~\ref{fig:sigma_vs_absz}.   

\subsection{\label{sec:sim_digitization}Digitization}
\begin{figure}
  \includegraphics[width=0.9\columnwidth]{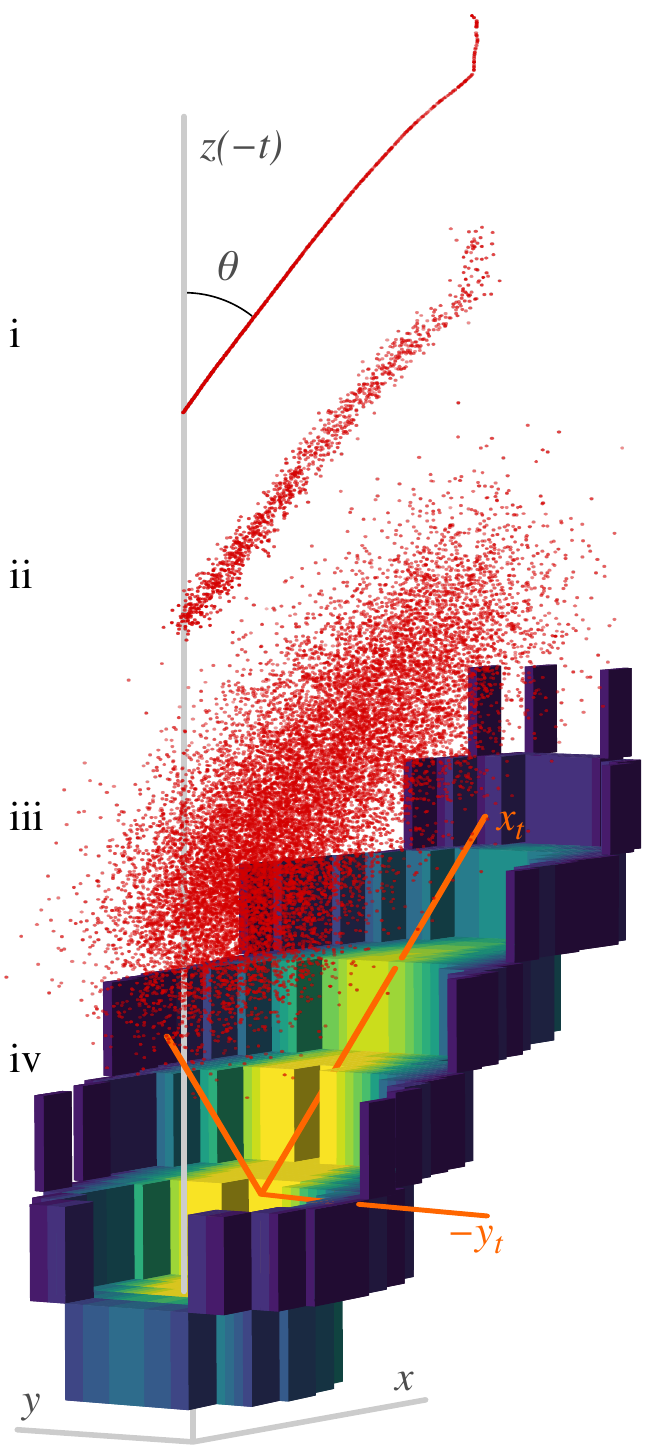}
  \caption{The drift and diffusion model, illustrated. The primary track (i) consists of electrons produced by ionization along the track of the recoiling nucleus (one red dot represents 10 electrons). After drift (ii), the primary electrons have diffused in $x-y$ and independently in $z$. After amplification (iii), the number of electrons has been multiplied by the gain (here reduced by a factor of 100), with additional dispersion in both $x-y$ and $z$. After digitization (iv), the 2D charge density (color/shade) is measured in voxels with $z$ position determined by the relative threshold-crossing time of each pixel. The chip coordinate system is shown in gray, and the track coordinate system in orange. The track vector points in the $x_t$ direction. The track shown is a \SI{400}{\keV} He recoil detected in a BEAST TPC.}
\label{fig:model}
\end{figure}

To simulate the digitization, we integrate charge in $250\times\SI{50}{\micro\meter\squared}$ bins, matching the size of the FE-I4b pixels. A four-bit TOT value in each bin is calculated from a chip calibration that maps injected charge to TOT. The relative timing in $z$ is based on digitizing the threshold-crossing time for each pixel in time bins determined by the drift velocity and clock speed, relative to the first threshold crossing. These bins are $\SI{250}{\micro\meter}$ in $z$, so the hits can be approximated as voxels with size $250\times50\times\SI{250}{\micro\meter\cubed}$. The resulting track then consists of a number of ``hits'', one per pixel over threshold, with $x$ and $y$ coordinates given by the center point of the hit pixel, $z$ coordinate assigned relative to the first threshold-crossing in the event, and with an associated TOT value. Such a ``2.5D'' track is visualized in Fig.~\ref{fig:model}.

\section{\label{sec:model}Physical model and simulation}
A nuclear recoil track develops in the detector over four distinct stages, illustrated in Fig.~\ref{fig:model}. In the sections below, one for each stage (and labeled as in the figure), we describe the physics involved and develop a simplified and continuous mathematical model for track development, amplification and digitization of the BEAST TPCs.

It is convenient to define two coordinate systems: the \textit{chip} coordinates $x, y,$ and $z$ are defined so that the readout plane is segmented in $x$ and $y$ and the drift is antiparallel to $z$. The \textit{track} coordinates $x_t, y_t,$ and $z_t$ are defined so that $\hat{x}_t$ is the track vector and points along the primary axis of the track, $\hat{y}_t$ is transverse to the track and strictly in the $x-y$ plane, and $\hat{z}_t$ is perpendicular to both $\hat{x}_t$ and $\hat{y}_t$, as shown in Fig.~\ref{fig:model}. For the model, we consider, without loss of generality, a track that begins at $(x,y,z)=(0,0,z_{\text{abs}})$ with a track vector pointing in the $x$ direction with an angle of $\theta$ with respect to the $z$ axis. 

\subsection{\label{sec:primary}(i) The primary track}
The \textit{primary track} consists of all electrons generated via ionization by a nuclear recoil, and is visualized in Fig.~\ref{fig:model}(i) for a single recoil. The primary track contains information related to the properties of the recoil. The trajectory (angles $\phi$ and $\theta$) of the recoil is preserved, particularly in the first ionizations before multiple elastic scattering has redirected the recoil significantly. The ionization energy $E$ can be derived from the charge $q$ multiplied by the work function $w$. We define the primary track length $L$ to be the difference between maximum and minimum extent of all primary electrons when projected on the $x_t$ axis. 

\begin{figure}
  \includegraphics[width=\columnwidth]{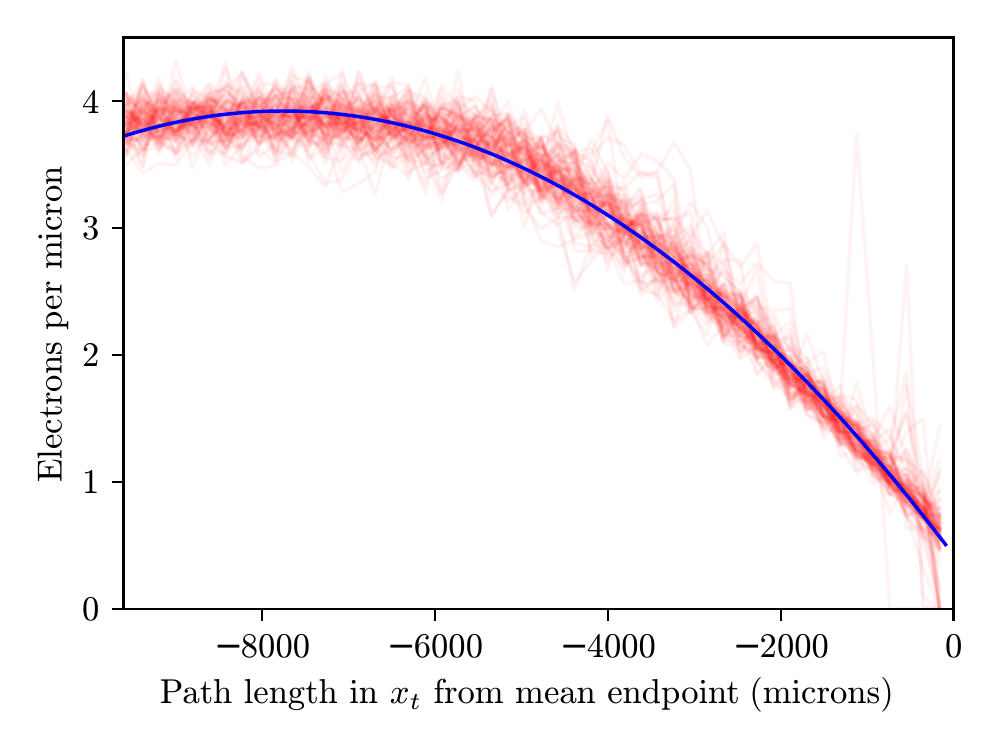}
  \caption{A parameterization of the Bragg curve (solid blue line) fitted to the linear charge density $\lambda$ of simulated alpha recoils (semi-transparent red lines), calculated as the number of electrons per micron in $\SI{192}{\micro\meter}$ bins for 100 $\SI{1}{\MeV}$ alpha recoils simulated with SRIM, projected onto the axis of the initial momentum vector of the recoil ($x_t$). The parameterization uses a second-order Chebyshev series, with $x_t=0$ set to the mean endpoint of all recoils.}
\label{fig:bragg}
\end{figure}

To model the charge distribution of the primary track, we consider the linear charge density projected on the $x_t$ axis $\lambda(x_t)$, which should follow the Bragg curve. For most recoils of interest ($<\SI{1}{\MeV}$), the entire track lies beyond the Bragg peak, and ionization density decreases rapidly with distance. The Bragg curve is non-analytic, therefore we aim to parameterize it via a fit to simulated recoils near the end of the recoil range. Due to the stochastic nature of ionization, this parameterization can only describe average behavior. To that end, we use a set of 100 alpha recoils from SRIM at 1~MeV to find the mean charge density in bins of $x_t$ (Fig.~\ref{fig:bragg}). For convenience, we fix the mean endpoint of the tracks to $x_t=0$. We use a Chebyshev series of order two to fit the data and model the linear charge density according to:
\begin{align}
  \lambda_{\textrm{B}}(x_t)=
  \begin{cases}
    aT_0(x_t)+bT_1(x_t)+cT_2(x_t) & \textrm{if } -L < x_t <0,\\ 
    0 & \textrm{otherwise}.
  \end{cases}\label{eq:bragg_parameterization}
\end{align}
For $x_t$ in microns and $\lambda$ in electrons per micron, we obtain parameter values of $a=0$, $b=\SI{-9.275e-4}{}$, and $c=\SI{-2.722e-8}{}$ from a least-squares fit to the alpha recoils in Fig.~\ref{fig:bragg}. An offset in $x_t$ will be required to align the endpoint of the parameterization with the endpoint of the track. We do not attempt to parameterize other recoil species, although in principle Eq.~\ref{eq:bragg_parameterization} can be reused or extended at will.

\subsection{\label{sec:drift}(ii) Drift and diffusion}
As the primary charges drift toward the GEMs, the charges diffuse independently in the \textit{transverse} plane ($x-y$) and in the \textit{longitudinal} direction ($z$), in an amount proportional to the square root of the drift distance $z_d$:
\begin{align}
  \sigma_T^{ii} = B_T \sqrt{z_d},\\
  \sigma_L^{ii} = B_L \sqrt{z_d},
\end{align}
where the constants $B_T$ and $B_L$ are determined by the gas and field properties and can be calibrated as in Ref. \cite{absz}, and the superscripts refer to the stage.

\subsection{\label{sec:amplification}(iii) Amplification}
As the track passes through the amplification stage, the number of electrons is multiplied by the effective gain $G$. Further broadening due to the GEM hole spacing and diffusion adds a fixed amplification resolution for all tracks, which we model using constants $A_T$ and $A_L$ as in Eqs.~\ref{eq:sigT} and \ref{eq:sigL}:
\begin{align}
  \sigma_T^{iii} = \sqrt{A_T^2 + B_T^2 z_d}\label{eq:sigmaT_vs_z},\\
  \sigma_L^{iii} = \sqrt{A_L^2 + B_L^2 z_d}\label{eq:sigmaL_vs_z}.
\end{align}

Rotation into the track coordinate system results in \textit{mixing} between the transverse and longitudinal diffusions (note that $\theta=90^{\circ}$ is a ``flat'' track parallel to the $x-y$ plane):
\begin{align}\label{eq:mixed_sigmas}
  \sigma_{x_t}^2 &= \sigma_T^2\sin^2\theta + \sigma_L^2\cos^2\theta,\\
  \sigma_{y_t}^2 &= \sigma_T^2,\\
  \sigma_{z_t}^2 &= \sigma_L^2\sin^2\theta + \sigma_T^2\cos^2\theta.
\end{align}
This simple form for $\sigma_{y_t}$ is the principal motivation for using the track coordinate system.

\subsection{\label{sec:digitization}(iv) Digitization}
After amplification, the drifting charge induces a signal on a segmented readout plane. For the purposes of this model, we assume that the signal develops instantaneously upon arrival of each electron. The validity of this assumption is discussed in Sec.~\ref{sec:demonstrations}. Each hit consists of the pixel coordinates $x^{h}$ and $y^{h}$, the threshold crossing time converted to a relative vertical position $z^{h}$, and the total integrated charge per pixel $q^{h}$, as visualized in Fig.~\ref{fig:model}(iv). The integration in $z$ leads to a loss of information: we measure two quantities per pixel ($z^h$ and $q^h$), while three are required to characterize the charge distribution in $z$ ($q^h$, $\mu_z$ and $\sigma_z$, assuming a Gaussian charge distribution). However, we will show that this information is recoverable by combining results from multiple pixels and by exploiting our charge distribution and digitization model. First, we write the charge distribution in $z$ before digitization:
\begin{align}
\frac{\text{d}q(z)}{\text{d}z} = q^{h} g_n(z; \mu_z, \sigma_z),
\end{align}
where $g_{n}(x; \mu, \sigma)$ is a Gaussian function with mean $\mu$ and width $\sigma$ and the $n$ indicates that it is normalized.

\begin{figure}
  \includegraphics[width=\columnwidth]{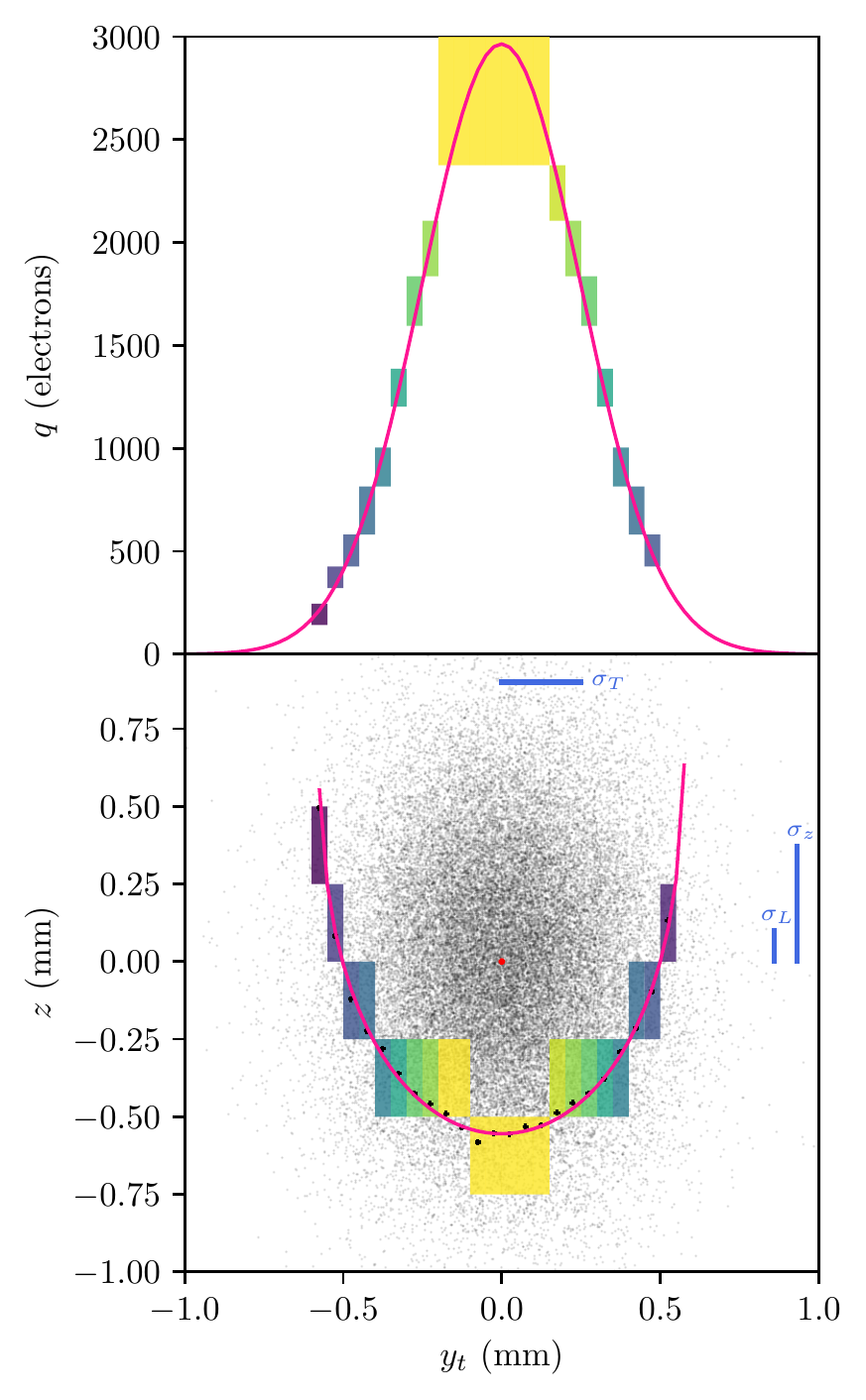}
  \caption{The charge profile (top) and shell (bottom) for a slice of a simulated charge cloud (gray dots) with $\theta=40^{\circ}$, $\sigma_T=\SI{250}{\micro\meter}$, and $\sigma_L=\SI{100}{\micro\meter}$. Charge is digitized using a pixel pitch of $\SI{50}{\micro\meter}$ in $y$, $\SI{250}{\micro\meter}$ in $z$, and with the charge scale (encoded by color/shade) from a FE-I4b calibration. The profile and shell fits, using Eq.~\ref{eq:qyt} and Eq.~\ref{eq:shell}, are shown with solid magenta lines. Due to the high inclination of the track, the dispersion in $z$ is dominated by the charge projection effect, as can be seen by comparing the bars representing $\sigma_L$ and $\sigma_z$ on the right of the bottom plot. The red dot in the center of the bottom plot marks the true track center $(\mu,z_0)$. The black points indicate the true threshold-crossing position in $z$ for each pixel before digitization.}
\label{fig:profile_shell}
\end{figure}

As the signal develops in a pixel, the charge is integrated in $z$, leading to a pulse shape that we represent with the error function $\mathrm{erf}$. For a given pulse with rise-time $\sigma_z$ and arrival time $\mu_z$, the threshold-crossing time $z^h$ depends on the pulse height $q^h$. We call this effect \textit{timecrawl}, as it is analogous to the well-known timewalk mechanism, but it is dependent on the charge-development time of the track, which is several orders of magnitude slower than the rise-time of the amplifier that determines timewalk. Due to the difference in charge density between the center and edges of a track, timecrawl is responsible for the distinct u-shaped transverse profile of the tracks, as illustrated in Fig.~\ref{fig:model}(iv). To parameterize this effect, we first note that at the moment of threshold-crossing, the integral is equal to the threshold charge $q_{\mathrm{th}}$ (where we write $z_0 = \mu_z$ to emphasize its interpretation as the ``true'' position of the track in $z$ for the charges in this hit):
\begin{align}\label{eq:qz}   
  q_{\mathrm{th}} &= \int_{-\infty}^{z^h}\text{d}q(z)  \nonumber \\
  &={q^h}\int_{-\infty}^{z^h}g_n(z; \mu_z=z_0, \sigma_z)\text{d}z \nonumber \\
  &= \frac{q^h}{2}\left[ 1+\mathrm{erf}\left(\frac{z^h-z_0}{\sqrt{2}{\sigma_z}} \right )\right ]. 
\end{align}
We now have two unknowns: $z_0$ and $\sigma_z$. To proceed, we now consider a \textit{slice}, which is to say a collection of hits transverse to the track (in $y_t$) for a narrow range of $x_t$. We expect the transverse charge profile to follow:
\begin{align}\label{eq:qyt}
  q^s(y_t) = h^s g(y_t;\mu_y^s,\sigma_T^s),
\end{align}
where the parameter $h^s$ contains information about the linear charge density at the slice, and the superscript $s$ indicates a quantity evaluated for an individual slice. We call this distribution the \textit{charge profile}, and it is illustrated in Fig.~\ref{fig:profile_shell}(top).

We can fit the charge profile to obtain values for $h^s$, $\mu_y^s$ and $\sigma_T^s$. Now, for a hit in the slice, we can replace $q^h$ with Eq.~\ref{eq:qyt} and rearrange Eq.~\ref{eq:qz} to express $z^h$ as a function of $y_t^h$ and the remaining parameters:
\begin{align}\label{eq:shell}
  z^h(y_t^h)=z_0+\sqrt{2}\sigma_z\textrm{erf}^{-1}\left[\frac{2q_{\mathrm{th}}}{h^s g(y_t^h;\mu_y,\sigma_T)} -1\right ].
\end{align}
This expression defines the shape of what we call the \textit{charge shell} for a transverse slice, as seen in Fig.~\ref{fig:profile_shell}(bottom). This shell can be fit using $z_0$ and $\sigma_z$ as free parameters.

\begin{figure}
  \includegraphics[width=\columnwidth]{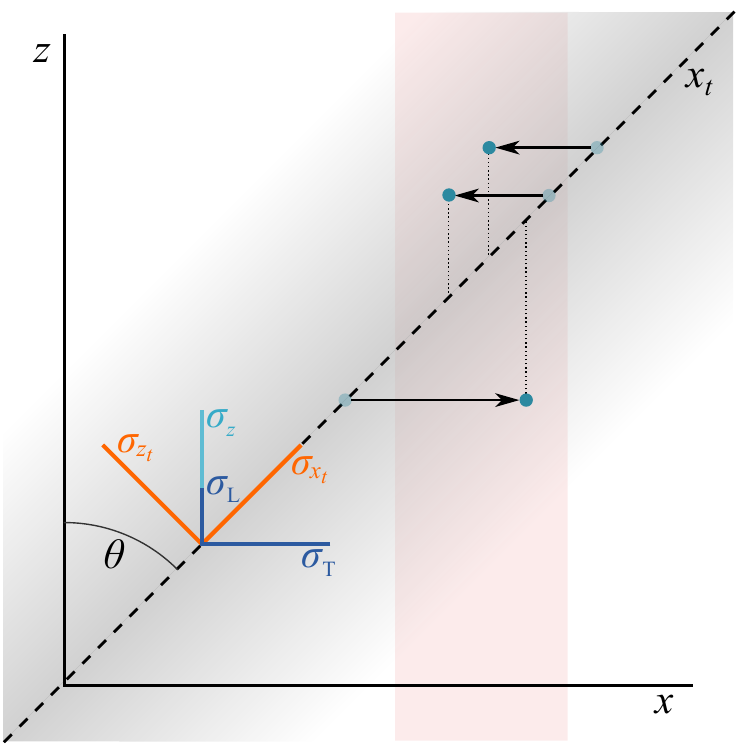}
  \caption{An illustration of the projection effect whereby transverse diffusion leads to smearing on the longitudinal axis (and vice versa) for inclined tracks.}
\label{fig:projection}
\end{figure}

To this point, we have shown that a charge profile fit ($q^h$ vs. $y^h$) combined with a charge shell fit ($z^h$ vs. $y^h$) can fully characterize the track in $z$ and $y$. Independently knowing $\mu_z$ and $\sigma_z$ allows us to measure $z_0$, which is the true position of the center of the track in $z$ relative to the first hit in the event and without diffusion. However, $\sigma_z$ is not simply equal to $\sigma_L$ for elevated tracks due to the \textit{charge projection effect} illustrated in Fig.~\ref{fig:projection}. Transverse diffusion from neighboring slices of the track causes charge to appear displaced from the track center in a given slice. Geometry then gives:
\begin{align}
  \sigma_z^2 = \sigma_L^2+\frac{\sigma_T^2}{\tan^2\theta}. 
\end{align}
Consequently, fits to the charge profile and shell determine the longitudinal diffusion $\sigma_L$, despite the loss of the charge structure information in $z$ due to integrating over time in each pixel. 

After the profile and shell fits, the values of $\sigma_T$ and $\sigma_L$ are known. Therefore we can write the linear charge density as a convolution:
\begin{align}\label{eq:smeared_bragg}
  \lambda(x_t) = \lambda_{\textrm{B}}(x_t;L,a,b,c)*g_n(x_t;0,\sigma_{x_t}),
\end{align}
using $\sigma_{x_t}$ from Eq.~\ref{eq:mixed_sigmas}. We call this the \textit{smeared Bragg} distribution, and the only free parameter is $L$.

\section{\label{sec:technique}Primary track recovery}
\begin{figure*}
  \includegraphics[width=0.93\textwidth]{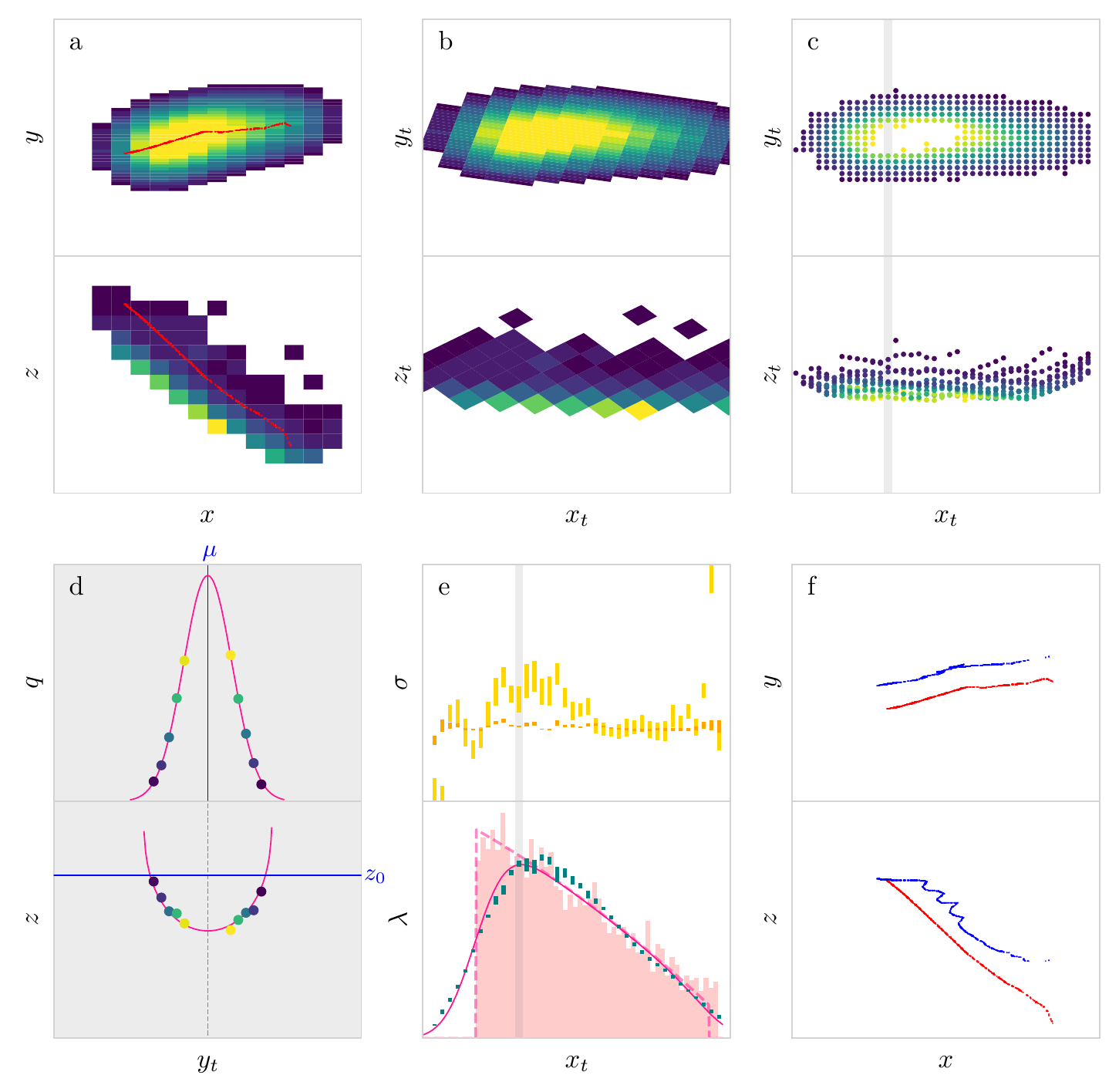}
  \centering
  \caption[short title]{\protect\begin{minipage}[t]{16cm}

    An illustration of the \ptr algorithm using one simulated 200~keV recoil. Panel labels (a, b, c...) correlate to section labels explaining each step. Each spatial dimension ($x, y, z, x_t, y_t, z_t$) is equal in span (roughly 1~cm). Colors indicate the primary track (red), fitted parameters (blue), and fitted functions (magenta). The color/intensity scale in all 2D histograms encodes charge density with lighter colors indicating higher densities. The panels show, in order:

    \begin{enumerate}[label=(\alph*)]
    \item The primary track (red dots) projected on the chip coordinates $x, y, z$, overlaid on top of the digitized signal.
    \item The digitized tracks rotated into the ``track'' coordinates $x_t, y_t, z_t$, obtained via the SVD prefit. 
    \item Samples (rows) organized by slice (columns). A bilinear interpolation estimates $q$ for each sample, shown by color/shade. The vertical gray bar indicates a single slice that is shown in panel d.
    \item (Top) the Gaussian fit to the transverse charge profile, along with the extracted value of $\mu$ (vertical solid blue line). (Bottom) the charge shell fit along with the extracted value of $z_0$ (horizontal solid blue line). 
    \item (Top) $\pm1$ sigma fit uncertainty bands for $\sigma_L$ (gold) and $\sigma_T$ (orange) for each slice. (Bottom) the smeared Bragg fit (solid magenta line), and the unsmeared Bragg component (dashed), fitted to the charge density (teal bars) of the slices. The pale red histogram in the background is the ``true'' simulated primary charge density projected on $x_t$ of the primary track.  
    \item Projections of the ``recovered'' (blue, above) and the true primary track (red, below), with arbitrary offsets in $y$ and $z$ for clarity.
    \end{enumerate}
    \end{minipage}
  }
  \label{fig:ptr_mc}
\end{figure*}

Now that we have described a model of track development and digitization, we are ready to build the \ptr algorithm. The steps of the \ptr algorithm are illustrated in Fig.~\ref{fig:ptr_mc} using one simulated recoil. Each step is described below in subsections with labels indicating the relevant panel in the figure. 

\subsection{\label{sec:technique_a}(a) Digitization}
The initial hit data is converted into 3D space-points $x$, $y$, and $z$, and charge $q$ per pixel, calculated from the TOT using a calibration. The track angles $\phi$ and $\theta$ can be seen in the 2D projections $y$ vs. $x$ and $z$ vs $x$, shown in Fig.~\ref{fig:ptr_mc}(a).

\subsection{\label{sec:technique_b}(b) Prefit}
We use a Singular Value Decomposition (SVD) to find the track vector $\hat{x}_t$. To guarantee that $\hat{y}_t$ lies in the $x-y$ plane, we define it as:
\begin{align}
  \hat{y}_t = \frac{\hat{z}\times\hat{x}_t}{\sin\theta},
\end{align}
where the factor $\sin\theta$ is necessary for normalization. We then define $\hat{z}_t=\hat{x}_t\times\hat{y}_t$. 

We call this decomposition the \textit{prefit} because the orientation of the track vector $\hat{x}_t$ in chip coordinates gives us $\theta$ and $\phi$, up to a head-tail flip $\hat{x}_t \to -\hat{x}_t$. We can also measure the track length $L$ (lengthened due to diffusion) and total detected ionization energy $q$ (with saturation and threshold losses) at this stage. We use these as benchmarks for evaluating improvement of the measurements of these variables with \ptr.

\subsection{\label{sec:technique_c}(c) Slicing and sampling}
We have written the track model in terms of transverse \textit{slices}. After the prefit, we define a series of slices transverse to the track, regularly spaced in $x_t$. For each slice, we define a number of \textit{samples} that are regularly spaced in $y_t$. For each sample, we obtain an estimate of the charge $q$ and coordinate $z$ via a bilinear interpolation from the four surrounding pixels. In order to avoid sample biases caused by edges or pixel saturation, we discard a sample if any one of these surrounding pixels either doesn't have a hit or is saturated. The resulting distribution of samples can be seen in Fig.~\ref{fig:ptr_mc}(c; top).

The slice and sample spacing should be comparable to or below the segmentation of the detector. For our studies, we use $\SI{100}{\micro\meter}$ spacing for both slices and samples, resulting in a regular grid of samples that slightly undersamples in $y$ (with a pixel pitch of $\SI{50}{\micro\meter}$) and oversamples in $x$ (pixel pitch $\SI{250}{\micro\meter}$). 

\subsection{\label{sec:technique_d}(d) Charge profile and shell fits}
For each slice, we first fit the charge profile, Fig.~\ref{fig:ptr_mc}(d; top), and then the charge shell (bottom), using the parameterizations shown in Section~\ref{sec:digitization}. The profile fit determines the parameters $h^s$, $\sigma_T^s$, and $\mu^s$. The shell fit determines the parameters $z_0^s$ and $\sigma_L^s$ by exploiting the timecrawl. We can interpret the coordinates $(\mu^s, z_0^s)$ as an update of the primary track position in $(x_t, z)$. These deviations are sensitive to deflections in the primary track due to multiple elastic scattering. From $h^s$ we can derive the linear charge density at the $x_t$ position of the slice:
\begin{align}
  \lambda^s(x_t) = \frac{h^s\sigma^s_T \sqrt{2\pi}\sin\theta}{G L_x L_y},
\end{align}
where the numerator contains the integral of the (unnormalized) Gaussian from the profile fit corrected for charge pileup in $\theta$, and the denominator includes the area of each pixel and the gain $G$.

In addition to the primary track properties, we also obtain values for $\sigma^s_L$ and $\sigma^s_T$ for each slice. Due to the low sensitivity for $\sigma_L^s$ from a single shell fit, the results are best averaged over a full track. In contrast, $\sigma_T^s$ is well-determined from a single slice and can in principle be used to measure the change in transverse diffusion between ends of tracks with a significant extent in $z$. The Gaussian width in the profile fit is insensitive to the detector threshold, bypassing a major source of width error in traditional techniques.

\subsection{\label{sec:technique_etop}(e; top) Calculation of track-level widths}
After completing all slice fits, we combine the results into track-level quantities. For $\sigma_L^{\ptr}$ and $\sigma_T^{\ptr}$, we take the weighted mean values over all slices, with weights equal to the inverse fit uncertainty squared. The resulting values are systematically broadened due to the resolution of the pixel chip, which for rectangular pixels depends on $\phi$. In order to generalize the resolution effects of a track with an arbitrary orientation to a rectangular grid, we define the \textit{effective pixel pitch} $\mathcal{L}$ as the pitch of a virtual 1D pixel that has the same resolution as the rectangular pixels projected on a given axis. For the track axes, this gives:
\begin{align}
  \mathcal{L}_{x_t}^2 = L_x^2\cos^2\phi + L_y^2\sin^2\phi,\\
  \mathcal{L}_{y_t}^2 = L_x^2\sin^2\phi + L_y^2\cos^2\phi.
\end{align}
Because $\sigma_T$ is the width in $y_t$, its resolution effects depend on $\mathcal{L}_{y_t}$, while the Bragg smearing width is in $x_t$ and therefore depends on $\mathcal{L}_{x_t}$. In order to use the results from the profile fits to fix $\sigma_{x_t}$ in the smeared Bragg fit, we need to account for the different width errors in $y_t$ and $x_t$.

There are two resolution effects that lead to errors in the Gaussian width measurements. The first is due to the uncertainty in the position of a hit. For a uniform charge distribution, the standard deviation in the position measurement calculated in the interval $[0,\mathcal{L}]$ gives the \textit{binary resolution} $\mathcal{L}/\sqrt{12}$. Although our charge distribution is not uniform, we find that this remains a good estimate of the uncertainty in the hit position for our typical widths and pixel pitches. However, a second effect arises from the nonuniform charge distribution: the center of each pixel is always further away from the track center than the mean position of charges collected by that pixel. This \textit{width bias} depends on $\mathcal{L}$ and $\sigma$ and cannot be easily derived analytically. However, we find with simulation that it is well described by $\mathcal{L}/N$, where $N$ is a numerical factor that depends on the pixel pitch and the typical values of $\sigma_T$. For the BEAST TPCs, we find that using $N=14$ results in a zero mean bias. We therefore correct the measured $\sigma_T$ by subtracting these resolution effects:
\begin{align}\label{eq:sigmaT_corrected}
\sigma^{\ptr}_T = \sqrt{\sigma_{y_t}^2-\frac{\mathcal{L}_{y_t}^2}{12}}-\frac{\mathcal{L}_{y_t}}{14},
\end{align}
where we have used $\sigma_{y_t}$ to indicate the transverse width measured by the profile fits.


\subsection{\label{sec:technique_ebottom}(e; bottom) Smeared Bragg fit}
For the linear charge density (figure part e; bottom), we expect to see the primary Bragg curve smeared by $\sigma_{x_t}$ (Eq.~\ref{eq:mixed_sigmas}). To obtain the transverse component of $\sigma_{x_t}$, we take the unbiased estimate $\sigma^{\ptr}_T$ obtained with Eq.~\ref{eq:sigmaT_corrected} and add the expected resolution biases in $x_t$:
\begin{align}
\sigma_T^2 = \left(\sigma_T^{\ptr}+\frac{\mathcal{L}_{x_t}}{14}\right)^2+ \frac{\mathcal{L}_{x_t}^2}{12}.
\end{align}\label{eq:sigmaT_corrected2}
We then use Eq.~\ref{eq:smeared_bragg} to fit this distribution, with an additional vertical scale factor $s$ (nominally $1$) that accommodates fluctuations in effective gain. The free parameters in the fit are the primary track length $L^{\ptr}$, defined as the offset between the end of the track and $x_t=0$, and $s$. The total charge $q^{\ptr}$ is then simply the analytic integral of the \textit{unsmeared} Bragg function (the dashed line in part e; bottom) with the length and scale factor determined from the fit.

The uncertainty assigned to the linear charge density for each slice comes from the parameter uncertainty in the charge profile fit. However, neighboring slices are correlated due to the bilinear interpolation. This leads to an artificially increased $\chi^2$ value, even for good fits. This can be seen in Fig.~\ref{fig:ptr_mc} (e; bottom), where modest deviations from the continuous model persist for multiple sequential slices. However, we see that the agreement between the unsmeared Bragg and the primary charge distribution is excellent. Therefore we do not attempt to decorrelate the slice errors, though in principle this could improve fit convergence performance in some cases.
 
We perform the smeared Bragg fit under two hypotheses: one in the default orientation (whatever $\phi$ and $\theta$ the SVD prefit gives us), and one in the ``flipped'' orientation (with $\hat{x}_t \to -\hat{x}_t$). We use the $\chi^2$ of the fits to select the best hypothesis; this gives us the \textit{head-tail} determination. 

\subsection{\label{sec:technique_f}(f) Primary track generation and fit}
Although the above steps are sufficient to measure all of the desired primary track variables, it may be useful to use these results to generate a reconstructed primary electron distribution. This can be used to compare features like path deviations that are not captured in the existing variables (which assume a straight track). To accomplish this, we generate $q^{\ptr}$ random numbers using the unsmeared Bragg from Sec.~\ref{sec:technique_ebottom} as a PDF. This gives us a distribution of charges with length $L^{\ptr}$ with the correct sense, as a function of $x_t$. For each charge, we add deviations in $z$ and $y_t$ based on a linear interpolation between the two surrounding slices of $z_0^{s}$ and $\mu^s$. Finally, the charges are rotated into the chip coordinate system using $\theta$ and $\phi$ from the prefit.

Fig.~\ref{fig:ptr_mc}(f) shows the 2D projections of the predicted primary track compared to the simulated primary track. We use the same SVD method as before to find $\theta^{\ptr}$ and $\phi^{\ptr}$ using this predicted primary track. Due to the larger number of charges at the beginning of the primary track, this final fit is effectively charge-weighted. Because the angle is best preserved in the beginning of the track, $\theta^{\ptr}$ and $\phi^{\ptr}$ may improve on the prefit determinations of the angles. However, as is evident in the figure, the $z^s_0$ at the slice level is not well determined, and the resulting value of $\theta^{\ptr}$ will be compromised. 

\subsection{\label{sec:technique_absz}Absolute $z$}
Although not visualized in Fig.~\ref{fig:ptr_mc}, \ptr also provides an estimate of \absz from $\sigma_T^{\text{ptr}}$. In principle, this could be done with $\sigma_L^{\text{ptr}}$ as well, but because the longitudinal width is more difficult to measure, it cannot constrain \absz as well.

In order to relate $\sigma_T$ to an absolute $z$ position, we need to know the diffusion parameters $A_T$ and $B_T$, which can be determined from simulation or by a calibration using sources with known positions in $z$~\cite{absz}. Once the diffusion parameters are known, \absz can be determined simply by:
\begin{align}\label{eq:absz}
  z_{\text{abs}} = \frac{\sigma_T^2-A_T^2}{B_T^2}.
\end{align}

The values of the diffusion parameters $A_T$ and $B_T$ are taken to be fixed inputs and therefore no uncertainty is propagated to \absz. Rather, the uncertainties can be determined experimentally, as in Sec.~\ref{sec:calibration}.

Due to the relationship $z_{\text{abs}}\propto\sigma_T^2$, the determination of \absz is very sensitive to the resolution of and biases in $\sigma_T$. In the square pixel case the calibration and extraction can both use biased widths without difficulty. However, in the rectangular pixel case, such as in the BEAST TPCs, it is necessary to control for variations in the width bias with $\phi$, as described in Sec.~\ref{sec:technique_etop}, in order to optimize the determination of \absz. 

\subsection{Summary of \ptr}
We have presented a technique for the effective deconvolution of all known detector effects, including diffusion and charge integration, from nuclear recoils. There are five key features of this algorithm. First, it provides a measurement of $\sigma_L$ and $\sigma_T$ at the track level, which can be used for 3D detector fiducialization. Second, once the diffusion is known, it can be deconvolved from the linear charge density distribution to provide a measurement of the primary track length $L$. Third, the Bragg fit allows us to measure the entire primary track charge $q$ without sensitivity to pixel saturation or threshold losses. Fourth, it provides head-tail sensitivity, determining the 3D vector directionality of the track. Fifth, the generation of the predicted primary track and subsequent determination of $\theta$ and $\phi$ is effectively charge-weighted and therefore should match the initial direction of the recoil better than the unweighted prefit. Therefore, we expect significant improvements in accuracy for $\theta$, $\phi$, $L$, and $q$.

We note a few additional capabilities of \ptr that may warrant further study. The $\chi^2$-based hypothesis testing of smeared Bragg fits can be used to test parameterizations of the Bragg curve for other nuclear species or electron recoils in the target gas. This could prove an efficient particle identification tool. The floating scale factor $s$ may be useful for gain monitoring or particle identification, or as a constraint to minimize the effects of statistical fluctuations. Additionally, \ptr has the ability to track the transverse straggling of a recoil due to multiple scattering. This can be used to ``unstraggle'' detected tracks, which may significantly improve the angle and head-tail determinations for the lowest energy tracks and could be particularly useful in reconstructing electron recoils.

In the following sections, we test the performance of \ptr on a large set of simulated recoils, and test aspects of it on TPC data with and without a neutron source.

\section{\label{sec:performance}Performance}
Before attempting data-driven validations of \ptr, we quantify the expected performance of its outputs with 1,100 simulated alpha recoils used to generate over one million (\num[group-separator={,}]{1306800}) tracks, using the simulated BEAST TPCs described in Sec.~\ref{sec:intro_simulation}. We compare to \textit{benchmark} methods that we have used previously~\cite{beast_tpcs}, which we will describe for each variable. 

We expect the performance of both the benchmark measurements and \ptr to depend primarily on the primary track energy $E^{\text{truth}}$ and polar angle $\theta^{\text{truth}}$. For low energies, the diffusion may be comparable to or larger than the track length, making the prefit inaccurate. For highly inclined tracks, the limited projection of the track on the readout plane can lead to a poor prefit and a significant loss of charge density information due to the compression of a large portion of the track onto a few pixels, as well as high levels of charge loss due to pixel saturation. Consequently, throughout this section we show measures of performance versus $E^{\text{truth}}$ and $\theta^{\text{truth}}$.

\begin{figure*}
\centering
\includegraphics[width=\textwidth]{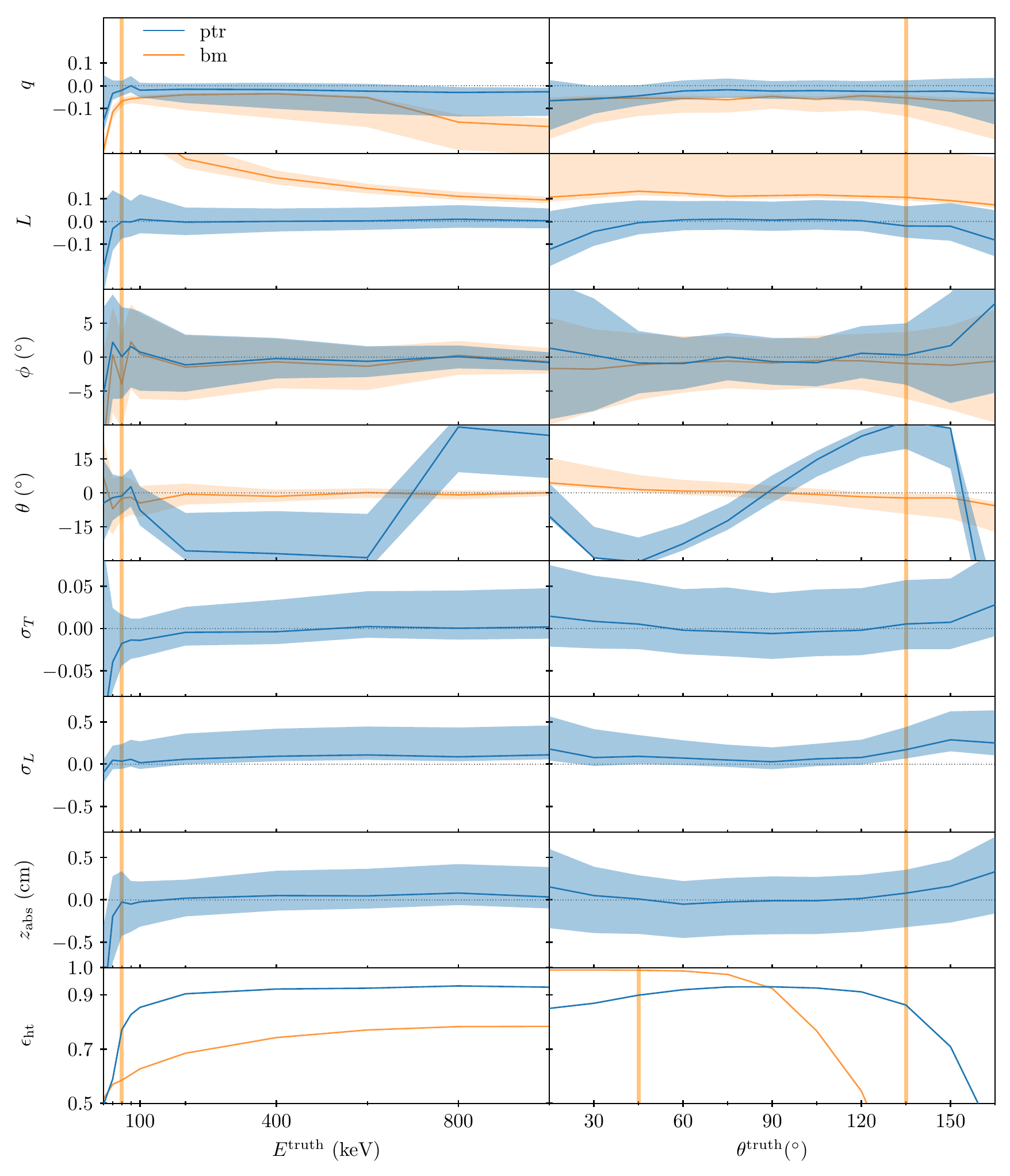}
\caption{A summary of the performance of \ptr versus the benchmark methods using the simulated dataset. Each row shows the bias (solid lines) and resolution (shaded regions) of the parameter errors versus truth energy $E^{\text{truth}}$ and polar angle $\theta^{\text{truth}}$. For the angles $\theta$ and $\phi$, errors due to head-tail misassignment have been removed. Parameter errors are normalized to the truth value of the parameter except when units are given, in which case the absolute errors are shown. The last row shows the mean head-tail efficiency $\epsilon_{\text{ht}}$ for \ptr versus the benchmark method. Vertical orange bars show the energy and polar angle of the standardized working point. }
\label{fig:bias_res}
\end{figure*}

Our primary performance metrics are \textit{bias} and \textit{resolution}. We define bias to be the most probable value (MPV) of the error. In order to account for strongly assymetric tails, for the resolution we use the standard deviation of the error calculated independently below and above the MPV. We show summaries of the bias and resolution for all key variables as functions of $E^{\text{truth}}$ and $\theta^{\text{truth}}$ for \ptr and the benchmark methods in Fig.~\ref{fig:bias_res}.

In order to facilitate extrapolation to other detectors, we additionally gauge performance at a \textit{standardized working point} (SWP) of $E^{\text{truth}}=60$~keV and $\theta^{\text{truth}}=135^{\circ}$. This corresponds to a mean \lst value of 2.15 (foreshortened to $2.15/\sqrt{2}=1.52$ on the readout plane), with significant but modest integration effects. A summary of the performance of \ptr compared to the benchmark method at the SWP is shown in Table~\ref{tab:swp_performance}. 

\subsection{\label{sec:performance_sample}Performance study sample}
The MC dataset consists first of primary tracks generated by SRIM, with 100 recoils at each of the following ``truth'' energies:
\begin{align}
  E^{\text{truth}}\in & \{10,20,40,60,80,100,\nonumber \\
  & 200,400,600,800,1000\}~\textrm{keV}.\nonumber
\end{align}
We then simulate drift, amplification, and digitization, as described in Sec.~\ref{sec:intro_simulation}, for each of these primary tracks for each combination of the following values:
\begin{align}
  \phi^{\text{truth}} \in &\{0, 15, 30, 45, 60, 75, 90, \nonumber \\
  &105, 120, 135, 150, 165\}^{\circ},\nonumber \\
  \theta^{\text{truth}} \in &\{15, 30, 45, 60, 75, 90, \nonumber \\
  &105, 120, 135, 150, 165\}^{\circ},\nonumber \\
  z^{\text{truth}} \in &\{1, 2, 3, 4, 5, 6, 7, 8, 9\}~\textrm{cm}, \nonumber
\end{align}
where $z^{\text{truth}}$ is the absolute $z$ position of the center of the primary track. The primary track is centered on the middle of the pixel chip in $x$ and $y$, and the diffusion parameters are fixed to the values obtained from Magboltz: $A_T=\SI{143.4}{\micro\meter}$, $A_L=\SI{97.4}{\micro\meter}$, $B_T=\SI{134.8}{\micro\meter/\sqrt{\centi\meter}}$, and $B_L=\SI{128.2}{\micro\meter/\sqrt{\centi\meter}}$.

  \begin{table}
    \renewcommand{\arraystretch}{1.8}
  \caption{Performance summary for the key primary track properties using simulated tracks at the standardized working point (SWP). The first column indicates the measured track property. The second column indicates the type of quantity shown: ``frac. err.'' is the fractional error $(x-x^{\text{truth}})/x^{\text{truth}}$ for a variable $x$, ``abs. err.'' is the absolute error $x-x^{\text{truth}}$, and ``abs.'' is simply $x$. The next columns show the bias (most probable value of the error) and resolution (single-sided standard deviation of the error) for \ptr and for the benchmark method (bm). We indicate quantities that are significantly better for a particular method with boldface. In the case of the head-tail efficiency $\epsilon_{\text{ht}}$, we additionally consider $\theta^{\textrm{truth}}=45^{\circ}$ tracks in order to balance the asymmetric effects of integration for upgoing and downgoing tracks.}
\label{tab:swp_performance}
\begin{tabular}{llcccc}
\hline\noalign{\smallskip}
Variable                & Type                     & \multicolumn{2}{c}{Bias}             & \multicolumn{2}{c}{Resolution}\\
{}                      & {}                       & \ptr               & bm              & \ptr             & bm \\
\noalign{\smallskip}\hline\noalign{\smallskip}
$q$                     & frac. err.              & $\mathbf{-0.008}$   & $-0.066$        & $^{+0.032}_{-0.023}$           & $\mathbf{^{+0.017}_{-0.021}}$ \\
$L$                     & frac. err.              & $\mathbf{-0.034}$   & $0.600$         & $\mathbf{^{+0.092}_{-0.050}}$  & $^{+0.119}_{-0.066}$\\
$\phi$                  & abs. err. ($^{\circ}$)  & $-4.5$              & $-5.5$          & $^{+8.4}_{-5.0}$               & $^{+7.8}_{-5.4}$  \\
$\theta$                & abs. err. ($^{\circ}$)  & $8.7$               & $-17.3$         & $^{+6.0}_{-8.0}$               & $^{+4.6}_{-5.2}$  \\ 
$\sigma_T$              & frac. err.              & $-0.017$            & $-$             & $^{+0.032}_{-0.027}$          & $-$ \\
$\sigma_L$              & frac. err.              & $0.07$              & $-$             & $^{+0.20}_{-0.09}$           & $-$ \\
\absz                   & abs. err. (cm)          & $-0.05$             & $-$             & $^{+0.35}_{-0.39}$           & $-$ \\
$\epsilon_{\text{ht}}$  & abs.                    & $\mathbf{0.78}$     & $0.51$          & $-$              & $-$      \\

\noalign{\smallskip}\hline
\end{tabular}
\end{table}

\subsection{\label{sec:basic_performance}Basic performance}
Due to a requirement that \ptr has at least one valid slice consisting of at least four samples, for a given TPC there is a minimum energy below which \ptr cannot run. With the BEAST TPCs, out of $118,800$ tracks simulated at $10$~keV, zero were successfully processed. In contrast, over $\SI{45}{\percent}$ of tracks in the next energy bin, $20$~keV, succeeded. The success rate was over $\SI{94}{\percent}$ at $60$~keV and $\SI{100}{\percent}$ at $80$~keV and above. Only tracks that succeed are considered in this section. 

For tracks with sufficient length to be processed, a small number have smeared Bragg fits that fail to converge, or converge pathologically. This typically can be seen in the unnormalized, reduced $\chi^2$ value of the fit. By hand inspection, we find that below $\chi^2/N_{\text{ndf}}=30$, the large majority of fits are of high quality, while above that threshold the number of failed fits increases. Over $86\%$ of the tracks in the simulation sample are considered ``good'' quality by this definition, and those that are not are predominently lower-energy, high-inclination tracks. Consequently, we expect the performance of the quantities determined by the smeared bragg fit ($L^{\text{ptr}}$, $q^{\text{ptr}}$, head-tail) to be relatively poor in this region. For the performance studies in this section we retain all tracks except those that completely fail, identified with $\chi^2/N_{\text{ndf}}>100$ or $q^{\ptr}>50000$.

Basic quality selections applied to all tracks are detailed in Sec.~\ref{sec:efficiencies}, with corresponding efficiencies for both the simulation and data samples.

\subsection{\label{sec:performance_E}Performance: primary track charge $q$}
\begin{figure*}
\centering
\begin{subfigure}{.5\textwidth}
  \centering
  \includegraphics[width=\linewidth]{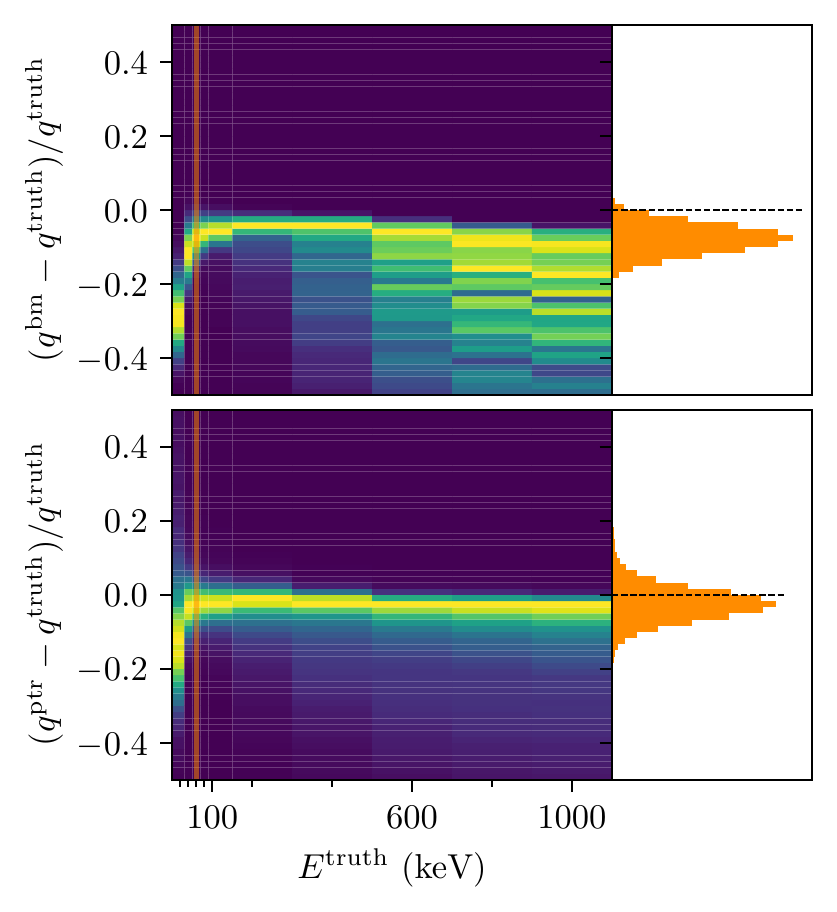}
  \label{fig:charge_vs_E}
\end{subfigure}%
\begin{subfigure}{.5\textwidth}
  \centering
  \includegraphics[width=\linewidth]{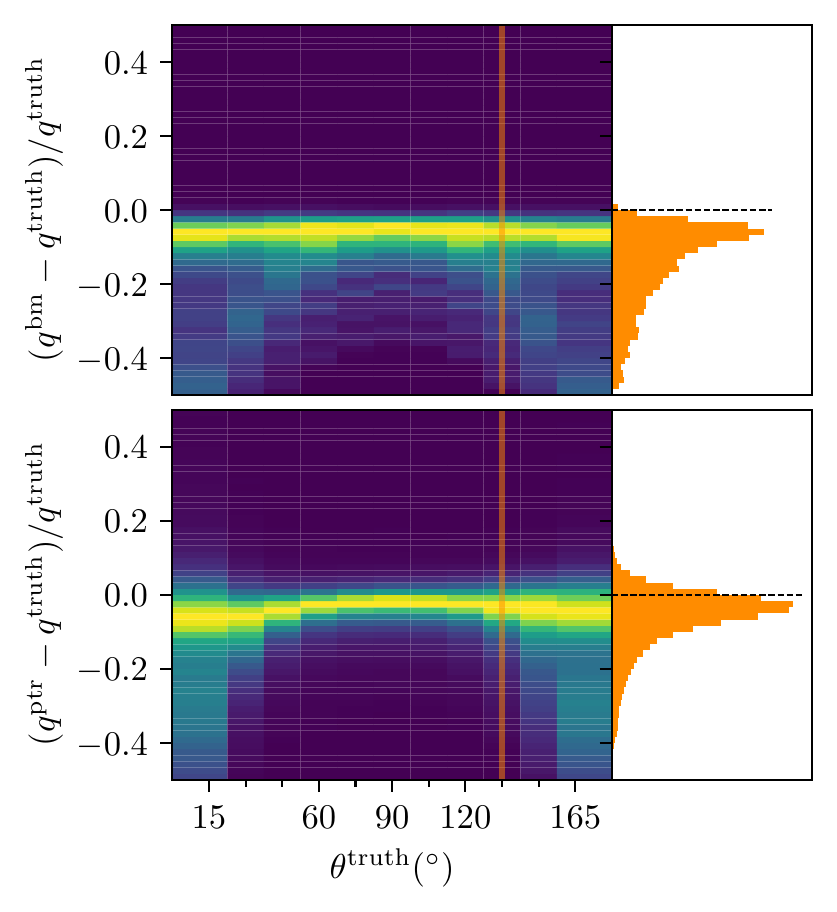}
  \label{fig:charge_vs_theta}
\end{subfigure}
\caption{The fractional error in the primary charge $q$ determination for benchmark (top) and \ptr (bottom) methods, versus primary track energy (left) and polar angle (right). The color/shade indicates the number of events in 2D bins, independently normalized in each column to the maximum value. Vertical orange bars show the energy and polar angle of the standardized working point and indicate the position of the profile histograms shown on the right.}
\label{fig:performance_charge}
\end{figure*}

The benchmark method of measuring the charge $q$ is to calculate the charge collected in each pixel and divide by the effective gain. There are two major sources of errors when calculating in this way, beyond uncertainty in the gain: first, pixels that do not cross threshold cannot be included in the sum. Second, charge over the maximum TOT value in a pixel also cannot be included in the sum. This loss due to saturation can be very large for tracks with short drift length or high inclination.

We expect \ptr to improve significantly on the benchmark method of determining primary track charge due primarily to the charge profile fit, which effectively recovers charge lost to threshold and saturation effects. Additionally, the total primary track charge comes from the integral of the unsmeared Bragg distribution, which enforces the statistical average $\text{d}E/\text{d}x$ for a track of a given length. 

Fig.~\ref{fig:performance_charge} shows the performance of the benchmark (``bm'') and \ptr determinations of the primary track charge $q$ versus $E^{\text{truth}}$ and $\theta^{\text{truth}}$, with bias and resolution summarized in Fig.~\ref{fig:bias_res}. We summarize the key features of these figures:
\begin{itemize}
\item For the benchmark method, charge lost below threshold dominates for very low energies, seen as a rapid drop-off below $100$~keV. At $40$~keV and above, \ptr recovers nearly all of this lost charge.
\item For the benchmark method, charge lost through saturation dominates for intermediate and high energies. This is responsible for the shift in the most probable value to lower energies, and for the broad tails in the negative direction, which are largely due to incresed saturation losses for highly inclined tracks. Evidently, \ptr successfully recovers most of the lost charge for tracks with minimal or intermediate inclinations, but cannot recover the lost charge for the most highly inclined tracks.
\item As expected, \ptr has minimal bias across all energies and inclinations. 
\item Low energies and high inclinations continue to pose a challenge for an accurate charge measurement due to integration effects, but \ptr expands the usable range.
\end{itemize}

Table~\ref{tab:swp_performance} shows the performance of \ptr and the benchmark method for $q$ at the SWP. We find that \ptr largely restores charge lost under threshold, but with a small resolution penalty. 

\subsection{\label{sec:performance_L}Performance: primary track length $L$}

\begin{figure*}
\centering
\begin{subfigure}{.5\textwidth}
  \centering
  \includegraphics[width=\linewidth]{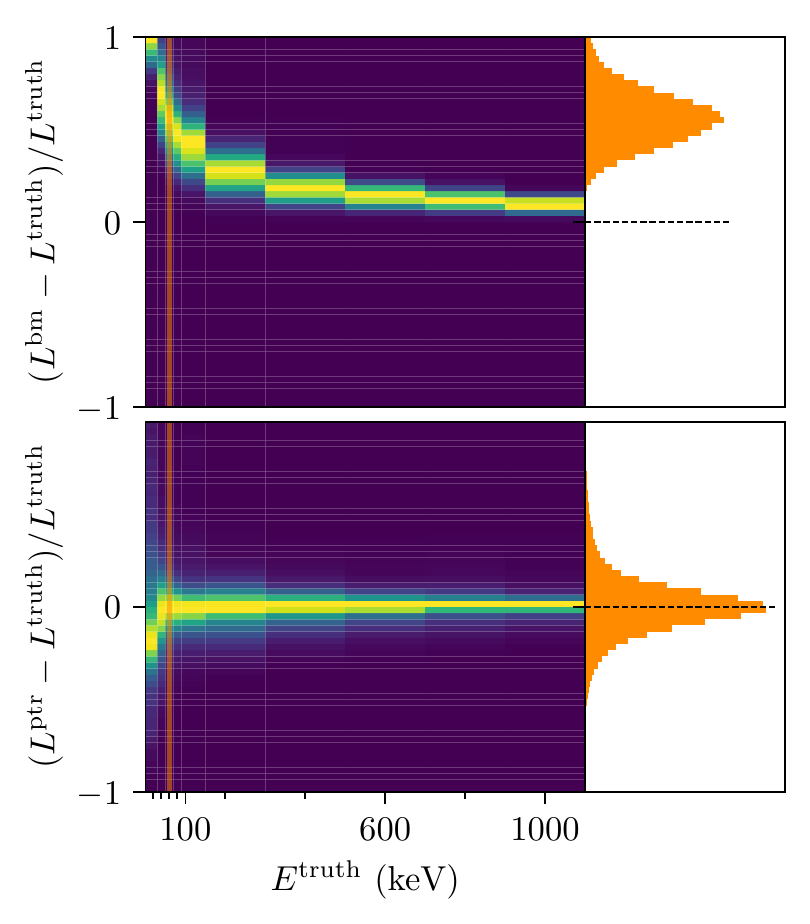}
  \label{fig:length_vs_E}
\end{subfigure}%
\begin{subfigure}{.5\textwidth}
  \centering
  \includegraphics[width=\linewidth]{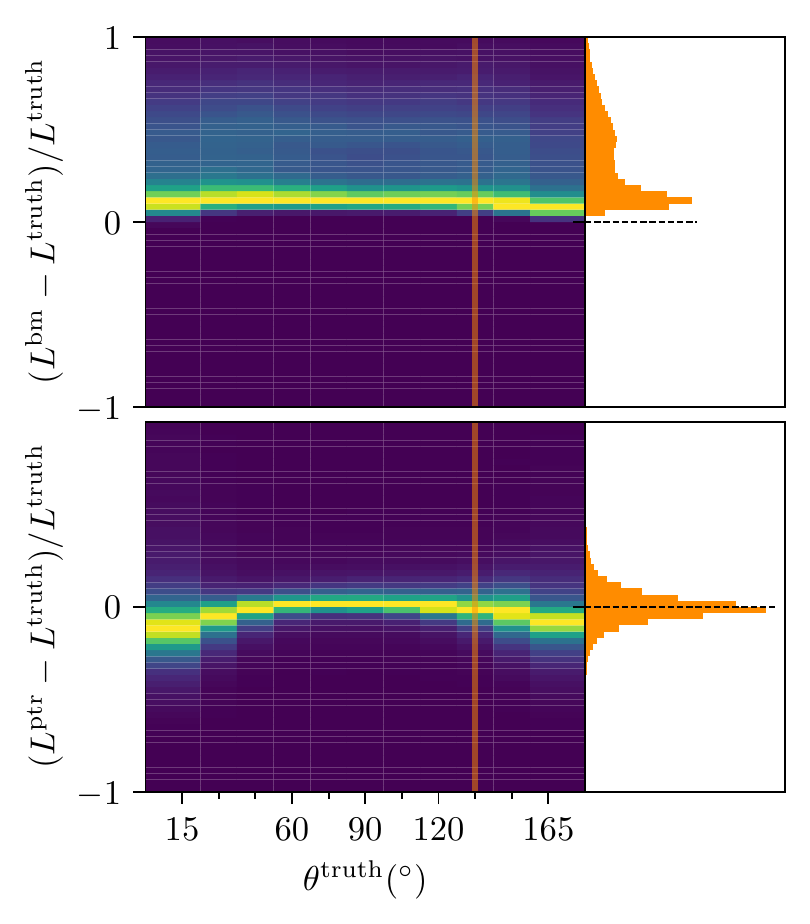}
  \label{fig:length_vs_theta}
\end{subfigure}
\caption{The fractional error in the primary length $L$ determination for benchmark (top) and \ptr (bottom) methods, versus primary track energy (left) and polar angle (right). The color/shade indicates the number of events in 2D bins, independently normalized in each column to the maximum value. Vertical orange bars show the energy and polar angle of the standardized working point and indicate the position of the profile histograms shown on the right.}
\label{fig:performance_length}
\end{figure*}

The primary track length $L$ can be used, in conjunction with energy, to separate different recoil species. The benchmark method is to simply take the length to be the difference between the maximum and minimum hit coordinates in $x_t$. We expect this method to always overestimate the track length due to diffusion broadening. However, we find that the lengthening is not simply $2\sigma_T$ due to the complex relationship between the spatial distribution of actual hits and variables like energy, width, and inclination. Additionally, in inclined tracks the trailing edges are missing hits due to charge integration (Fig.~\ref{fig:hcf_effect}), leading to $\theta$-dependent length errors independent of diffusion. 

In \ptr the length is a free parameter in the smeared Bragg fit, via the floating endpoints. Missing hits in the trailing edge of inclined tracks do not bias the fits, as the endpoints are free to travel beyond the ends of the distribution. Consequently, we expect \ptr to significantly improve the determination of the track length, primarily by removing the bias due to diffusion.

Fig.~\ref{fig:performance_length} shows the performance of the benchmark and \ptr determinations of the primary track length $L$ versus $E^{\text{truth}}$ and $\theta^{\text{truth}}$, with bias and resolution summarized in Fig.~\ref{fig:bias_res}. We summarize the key features of these figures:
\begin{itemize}
\item For the benchmark method, the length bias is large for low-energy tracks due to inclusion of the diffusion. At $40$~keV and above, \ptr naturally has no such bias.
\item The resolution of \ptr is somewhat worse than the benchmark at medium and high energies, perhaps as a result of range straggling, which is not accommodated in the smeared Bragg fit. 
\item For highly inclined tracks, \ptr slightly underestimates the length due to the the difficulty of performing the smeared Bragg fit in these cases. 
\end{itemize}

Table~\ref{tab:swp_performance} shows the performance of \ptr and the benchmark method for $L$ at the SWP. We find that \ptr corrects the length bias due to diffusion and improves the length resolution.

\subsection{\label{sec:performance_phi}Performance: primary track angles $\phi$ and $\theta$}

The benchmark method of measuring the track angles $\phi$ and $\theta$ is to derive them from the track vector obtained by the prefit. As discussed in Sec.~\ref{sec:intro_angles}, we do not expect to be able to significantly improve on the angular resolution of the benchmark method algorithmically. However, because \ptr's postfit is effectively charge-weighted, it may decrease errors caused by transverse straggling. Indeed, in Fig.~\ref{fig:bias_res} we see that the resolution for the azimuthal angle $\phi$ with head-tail swaps removed and at energies above $\SI{100}{\keV}$ is somewhat better for \ptr compared to the benchmark. However, the performance of \ptr for $\theta$ is far worse than the benchmark method due to the difficulty of extracting the track center coordinate $z_0$ in the shell fit. Consequently, we recommend using the benchmark polar angle $\theta$, but flipped according to \ptr's head-tail hypothesis test. 

\subsection{\label{sec:performance_sigmaT}Performance: widths $\sigma_T$ and $\sigma_L$}
The widths $\sigma_T$ and $\sigma_L$ are not primary track variables, but can be used in a number of applications, such as determination of the absolute $z$ position \absz and measurement of gas properties. Via the charge profile and shell fits, \ptr provides an estimation both of \absz derived from the transverse width $\sigma_T$ and the longitudinal width $\sigma_L$. The transverse width is not measured in the prefit, and the longitudinal width has never been measured on the track level for nuclear recoils, so we do not compare to benchmark values. 

\begin{figure*}
\centering
\begin{subfigure}{.5\textwidth}
  \centering
  \includegraphics[width=\linewidth]{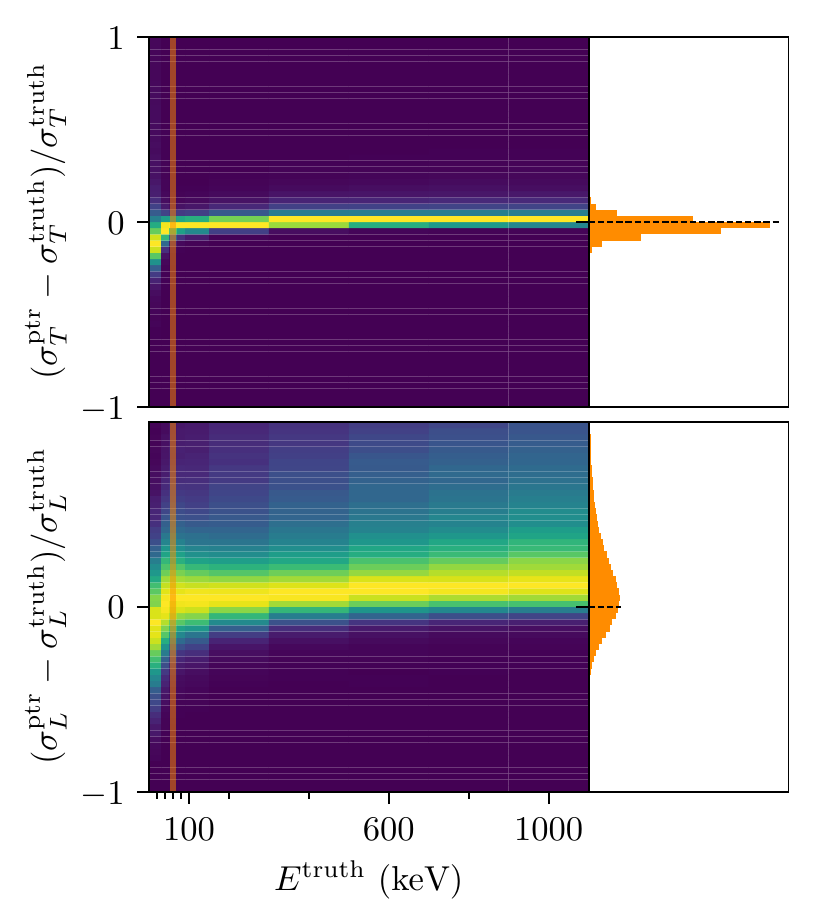}
  \label{fig:sigma_vs_E}
\end{subfigure}%
\begin{subfigure}{.5\textwidth}
  \centering
  \includegraphics[width=\linewidth]{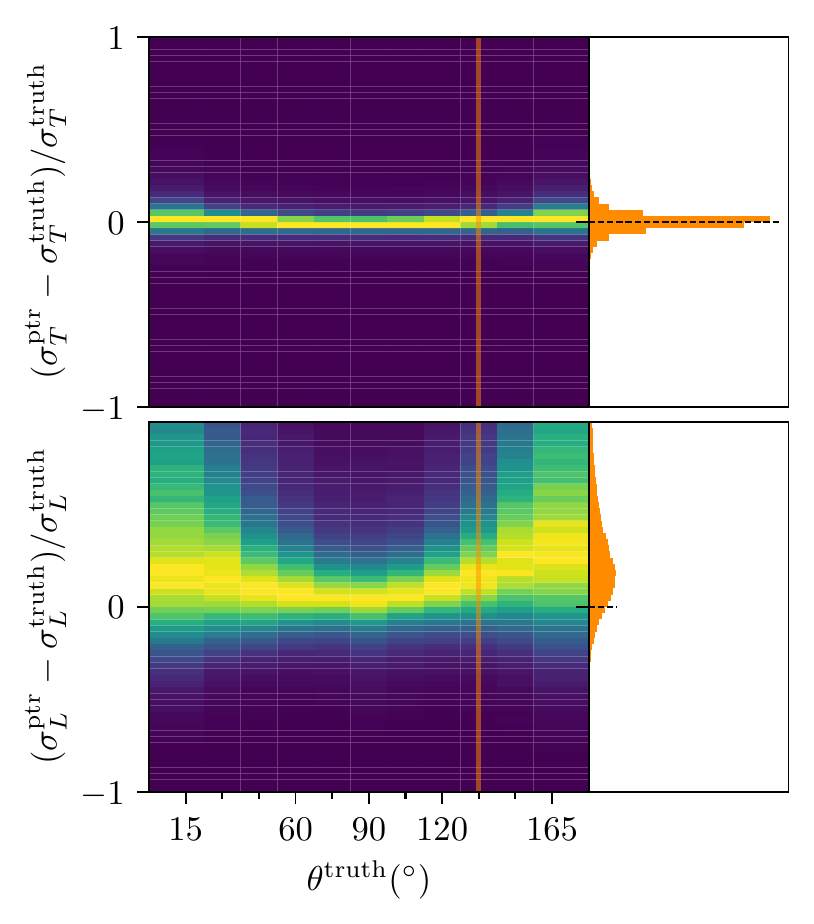}
  \label{fig:sigma_vs_theta}
\end{subfigure}
\caption{The fractional error in the widths $\sigma_T$ (top) and $\sigma_L$ (bottom), versus primary track energy (left) and polar angle (right). The color/shade indicates the number of events in 2D bins, independently normalized in each column to the maximum value. Vertical orange bars show the energy and polar angle of the standardized working point and indicate the position of the profile histograms shown on the right.}
\label{fig:performance_sigma}
\end{figure*}

Fig.~\ref{fig:performance_sigma} shows the transverse and longitudinal widths as measured by \ptr vs. energy and polar angle, with bias and resolution summarized in Fig.~\ref{fig:bias_res}. We summarize the key features of these figures:
\begin{itemize}
\item The transverse width $\sigma_T$ is determined with excellent resolution and minimal bias, down to the minimum energies.
\item The longitudinal width $\sigma_L$ has minimal bias across the full energy range, but with a large tail extending to significantly higher values than truth. These tails are due almost entirely to highly inclined tracks.
\end{itemize}

The \ptr model for the shell fit includes corrections for inclined tracks that we have validated with dedicated simulation. However, fitting highly inclined tracks still presents a substantial challenge due to integration effects. 

Table~\ref{tab:swp_performance} shows the performance of \ptr for $\sigma_T$ and $\sigma_L$ at the SWP. The determination of $\sigma_T$ is excellent, confirming that the resolution effects are properly handled. While the determination of $\sigma_L$ is not as good, we confirm that \ptr measures $\sigma_L$ despite the confounding effects of inclination and charge integration.

\subsection{\label{sec:performance_absz}Performance: absolute position $z_{\textrm{abs}}$}
One of the principal goals of \ptr is to provide a track-level determination of the absolute $z$ position via transverse width. This is possible by relating the measured transverse width $\sigma_{\ptr}$ to \absz via a calibration, as in Eq.~\ref{eq:absz}. For our performance studies, we use the same calibration constants $A_T$ and $B_T$ to derive \absz with \ptr as we do in the simulation. In practice, additional systematic errors will arise depending on the quality of the calibration. 

\begin{figure*}
\centering
\begin{subfigure}{.5\textwidth}
  \centering
  \includegraphics[width=\linewidth]{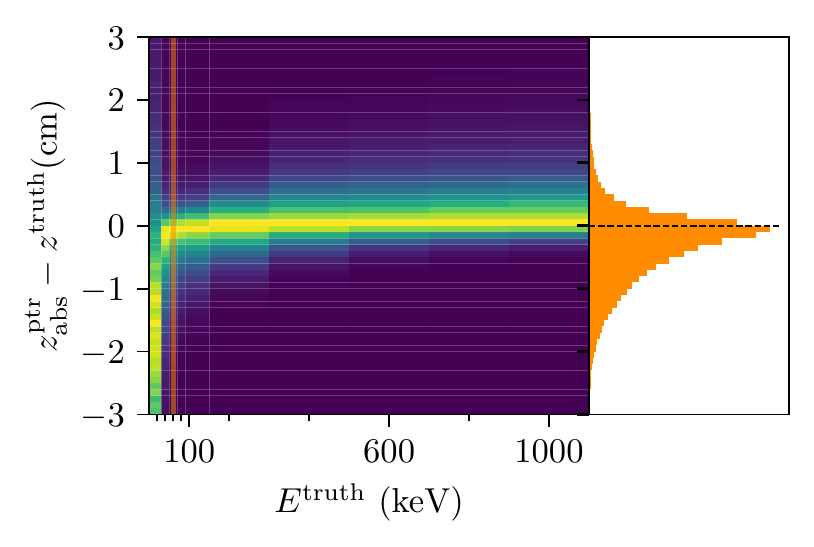}
  \label{fig:phi_vs_E}
\end{subfigure}%
\begin{subfigure}{.5\textwidth}
  \centering
  \includegraphics[width=\linewidth]{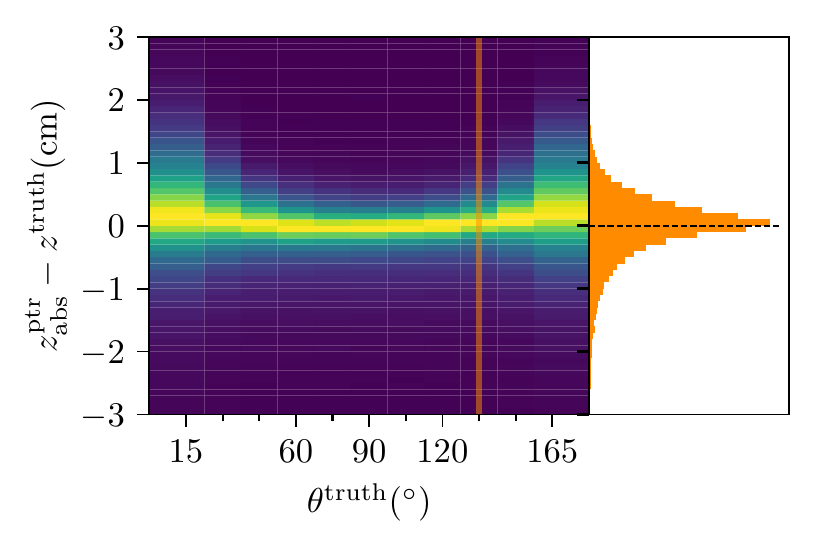}
  \label{fig:phi_vs_theta}
\end{subfigure}
\caption{The absolute error in the absolute position $z_{\text{abs}}^{\ptr}$ determination versus primary track energy (left) and polar angle (right). The color/shade indicates the number of events in 2D bins, independently normalized in each column to the maximum value. Vertical orange bars show the energy and polar angle of the standardized working point and indicate the position of the profile histograms shown on the right.}
\label{fig:performance_absz}
\end{figure*}

Fig.~\ref{fig:performance_absz} shows the performance of the \ptr determination of the absolute position \absz versus $E^{\text{truth}}$ and $\theta^{\text{truth}}$, with bias and resolution summarized in Fig.~\ref{fig:bias_res}. We summarize the key features of these figures:
\begin{itemize}
  \item Across a broad range of energies, \ptr successfully predicts the absolute position of the track on the drift axis. 
  \item For very low energies ($E^{\text{truth}}<40$~keV), $z_{\text{abs}}$ is pulled systematically down. This is due to insufficient profile information to constrain the Gaussian charge profile fit.
  \item Overall, the resolution in \absz is poor compared to the transverse width $\sigma_T$ due to the relatively shallow slope of \absz as a function of $\sigma_T$. This is a consequence of the amplification width in Fig.~\ref{fig:sigma_vs_absz}. 
\end{itemize}

Table~\ref{tab:swp_performance} shows the performance of \ptr for \absz at the SWP. We find that \ptr is capable of determining \absz even for this marginal case, but a $2\sigma$ ($>97\%$) rejection of cathode tracks would require a reduction in the fiducial volume of roughly $20\%$. Due to the sensitivity of the \absz determination to $\sigma_T$, we find that the resolution corrections shown in Eq.~\ref{eq:sigmaT_corrected} are essential for obtaining reliable and unbiased estimates of \absz independent of azimuthal angle $\phi$. 

\begin{figure}
  \includegraphics[width=\columnwidth]{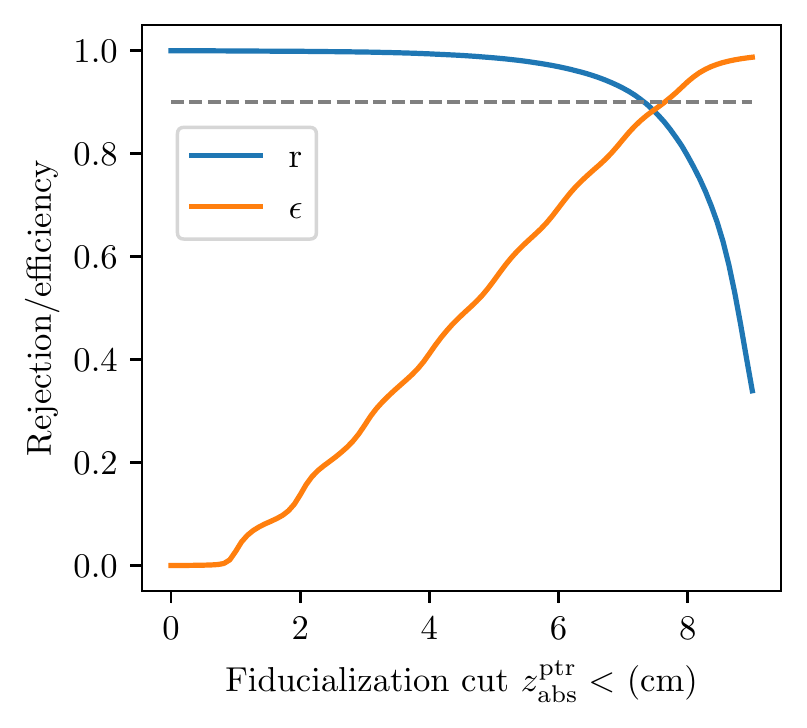}
  \caption{Cathode track rejection rate $r$ and signal efficiency $\epsilon$ versus fiducialization cut on $z_{\text{abs}}^{\ptr}$ on simulated tracks. For the purposes of this demonstration, we consider tracks with $z^{\text{truth}}=9$~cm to be ``background'' and all other tracks to be ``signal''. The dashed gray line indicates a rejection rate or efficiency of $0.9$.}
\label{fig:performance_fiducialization}
\end{figure}

The key metrics for evaluating fiducialization performance are the background rejection rate and signal efficiency as functions of the fiducialization cut. To approximate this for a BEAST TPC with 9~cm drift length, we show in Fig.~\ref{fig:performance_fiducialization} the fraction $r$ of $z^{\text{truth}}=9$~cm tracks rejected (as a proxy for cathode-emitted backgrounds) and $z^{\text{truth}}\in\{1,2,3,4,5,6,7,8\}$~cm tracks retained $\epsilon$ (as a proxy for signal) versus a fiducialization cut on $z_{\text{abs}}^{\ptr}$. Based on simulation alone, we expect that a cathode track rejection rate of $0.9$ is possible while maintaining a signal efficiency of nearly $0.9$. 

In the subsequent section we will test this fiducialization using data collected with the BEAST TPCs. Additionally, we will discuss design options that can improve the performance of the fiducialization for large-scale TPCs. 

\subsection{\label{sec:performance_headtail}Performance: head-tail efficiency}
To measure head-tail performance we define the \textit{head-tail efficiency} $\epsilon_{\textrm{ht}}$ as the fraction of tracks that are assigned the correct head-tail. We consider a head-tail assignment to be correct if the scalar product of the truth track vector and the reconstructed track vector has a value greater than zero. Since the prefit cannot determine head-tail its mean efficiency is $0.5$. For the benchmark method we use the \textit{head charge fraction} (HCF) of the track. A track is split into two halves at the midpoint between the maximum and minimum hit coordinate in $x_t$. The fraction of total charge contained in the half assigned as the head by the prefit is the HCF. If HCF$<0.5$, the track is assumed to have the correct head-tail assignment; if not, its track vector flipped.

\begin{figure}
  \includegraphics[width=\columnwidth]{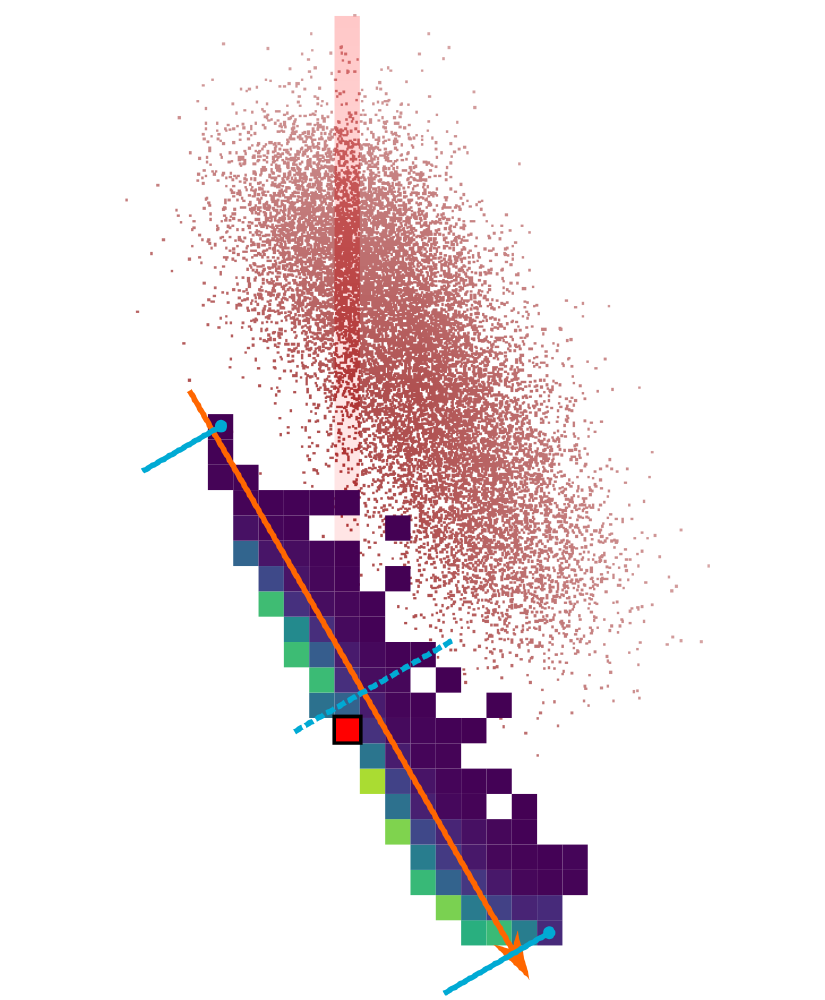}
  \caption{An illustration of integration effects in the head-charge fraction (HCF) calculation. Amplified charges (red dots) and the digitized track (2D histogram) are shown in an $x-z$ projection for a $400$~keV downgoing track with $\theta=150^{\circ}$. The majority of the charge is in the upper half of the charge cloud (further from the readout plane). However, when the detected track is split into two halves (blue lines), the lower half includes significantly more charge. To illustrate why, a single hit is highlighted in red with a black border. This hit contains all the charge integrated in a rectangular column indicated with pale red. Although the hit is assigned to the lower half based on its position relative to the track endpoints, most of its charge comes from the upper half of the charge cloud. This track is therefore assigned the incorrect head-tail using the HCF method. }
\label{fig:hcf_effect}
\end{figure}

We have recently used HCF~\cite{hedges} to determine head-tail for nuclear recoils using experimental data. However, we find here that charge integration effects significantly degrade its performance, particularly for highly inclined tracks. This effect is illustrated in Fig.~\ref{fig:hcf_effect}. Pixels in the lower half of the track integrate charge that belongs to the upper half, while the reverse isn't true. The same effect occurs in upgoing tracks, but in that case the effect amplifies rather than inverts the natural charge imbalance in the halves. We expect \ptr to be largely resistent to this effect up to high inclinations due to the floating endpoints that allow the unsmeared Bragg distribution to be shifted arbitrarily in $x_t$ with respect to the detected hits. 

\begin{figure*}
\centering
\begin{subfigure}{.5\textwidth}
  \centering
  \includegraphics[width=\linewidth]{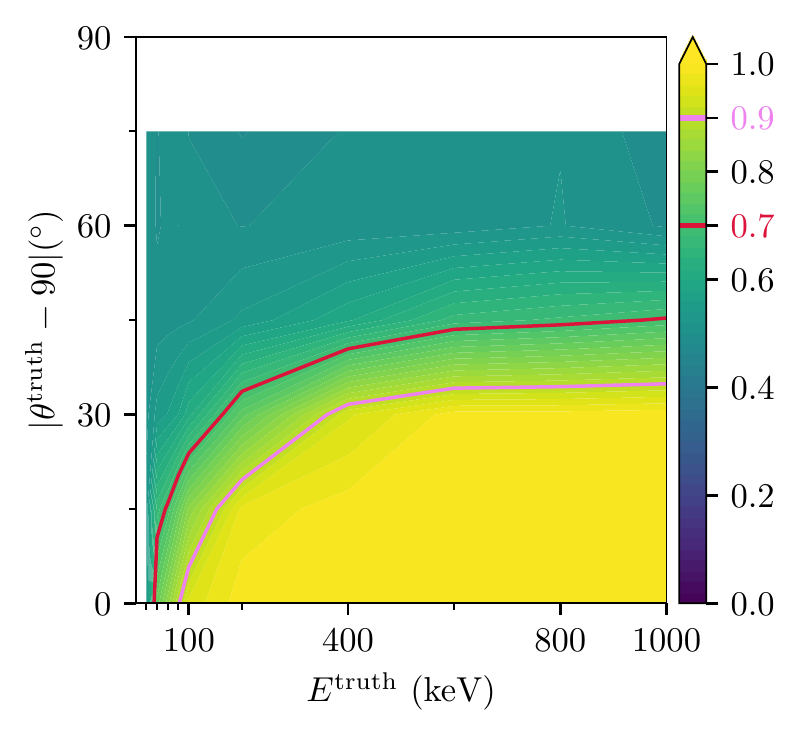}
  \label{fig:ht_contours_hcf}
\end{subfigure}%
\begin{subfigure}{.5\textwidth}
  \centering
  \includegraphics[width=\linewidth]{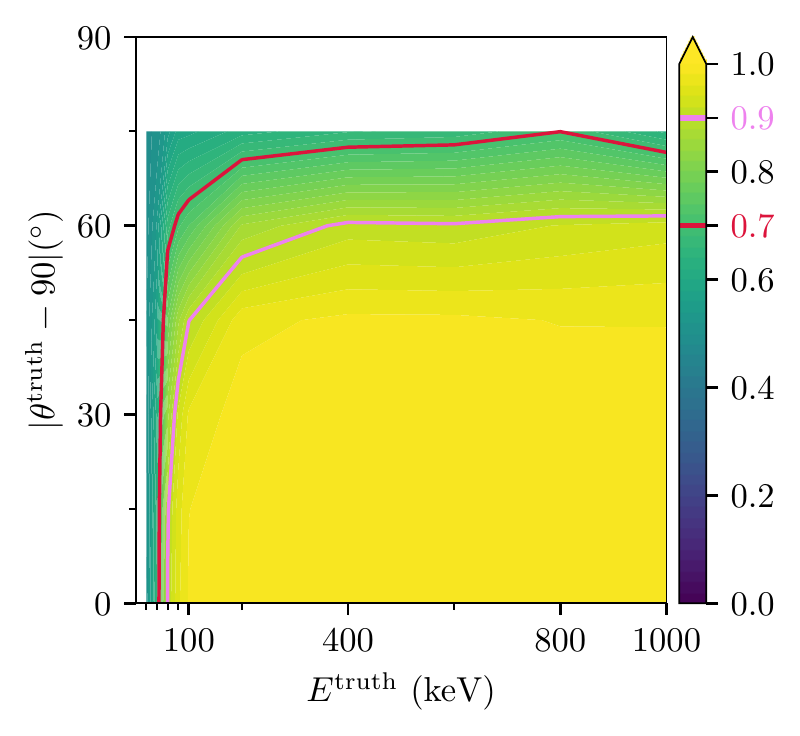}
  \label{fig:ht_contours_flip}
\end{subfigure}
\caption{Head-tail efficiency contours for HCF (left) and \ptr (right) in the inclination ($|\theta^{\text{truth}}-90^{\circ}|$) vs. energy ($E^{\text{truth}}$) plane. The solid contours are spaced every $2\%$. The red and lavender lines demarcate regions with head-tail efficiencies over $0.7$ and $0.9$, respectively.}
\label{fig:ht_contours}
\end{figure*}

The impact of integration effects on HCF can be seen in Fig.~\ref{fig:bias_res}. The apparent nearly perfect head-tail efficiency for $\theta<60^{\circ}$ is illusory, as in this region integration effects bias HCF in the direction of the correct head-tail assignment. In contrast, for $\theta>120^{\circ}$, the head-tail efficiency is nearly 0, far worse than the coin-toss efficiency for no head-tail determination at all. It is more appropriate to evaluate the head-tail efficiency as a function of the track inclination $|\theta^{\text{truth}}-90|$, which we show in Fig.~\ref{fig:ht_contours}. We find that both methods determine head-tail to very high efficiency for sufficiently high-energy tracks with low inclination. However, \ptr significantly outperforms HCF at medium to high inclinations due to better handling of charge integration effects. 

We additionally find the \ptr outperforms HCF at low energies. This can be seen at the SWP (Table~\ref{tab:swp_performance}). We also observe a similar improvement for tracks with no inclination, suggesting that the advantages of \ptr over HCF are not limited to controlling for integration effects. 

\subsection{Summary of performance}
We have demonstrated, using simulation, that \ptr improves on existing methods in the following ways:
\begin{itemize}
\item Bypassing diffusion and digitization effects in determination of primary track charge and length.
\item Proper handling of geometric effects such as charge projection. 
\item Providing a measurement of the longitudinal width on a track-by-track basis. 
\item Determining \absz with precision sufficient for detector fiducialization.
\item Resolution of head-tail ambiguity with high efficiency.
\end{itemize}
We find that \ptr can accomplish all of these things at or below the target working point of \lst$=2$, indicating successful deconvolution of physics and detector effects near the fundamental low-energy limit for any HD detector. However, this demonstration only proves that \ptr's modeling matches the simulated physics and digitization effects. It is still possible that the apparent performance improvements achieved by \ptr will not be seen on data, which would suggest shared deficiencies in both \ptr's models and in the simulation. To validate the performance, in the next section we use \ptr to recover primary track properties using data collected by the BEAST TPCs.

\section{\label{sec:demonstrations}Preliminary demonstrations with experimental data}

A comprehensive experimental validation of all products of \ptr will require independent dedicated testbeams in the future, preferably using a monochromatic neutron source. However, we can perform preliminary or qualitative tests of the primary capabilities of \ptr using existing experimental data from a system of 8 BEAST TPCs installed as beam background monitors for SuperKEKB integration~\cite{hedges,phase1,schueler}.

For $z_{\text{abs}}$ and fiducialization tests, we use the \textit{beam-off} dataset, which consists of over 120 TPC-days of recoil candidates collected by 8 TPCs at SuperKEKB with no beam, and therefore no neutron source except for naturally occurring radioactivity. For charge, length, and head-tail tests, we use the \textit{beam-on} dataset~\cite{schueler}, which consists of over 30 TPC-days of recoil candidates collected by a single BEAST TPC while SuperKEKB circulated beams. For a detailed technical description of the BEAST TPCs, including calibrations and resolution studies, see Ref.~\cite{beast_tpcs}.

To qualify as a neutron recoil candidate, a track must pass the following selections (hand-chosen):
\begin{itemize}
\item There are no hits on the edges of the chip (so that the entire track is visible).
\item The number of hits is greater than 100 (to reject most electron recoils).
\item The average charge per hit is greater than 7000 electrons (to reject most proton recoils). 
\end{itemize}
The primary purpose of these selections is to ensure that we have a highly pure sample of neutron recoils. We estimate, based on the number of $\textrm{d}E/\textrm{d}x$ outliers seen on experimental data (Fig.~\ref{fig:L_vs_q}), that the purity is in excess of $0.99$, which is sufficient for this preliminary demonstration.

After selections, the beam-off dataset contains 259 neutron recoil candidates, and the beam-on dataset contains over 230,000. The recoil energy spectra of both datasets follows a decaying exponential, with a mean value of $828(224)$~keV for the beam-off (beam-on) dataset. Due to the strong energy dependence in the performance of \ptr, we expect some differences in mean performance between the two datasets and simulation. Although the neutron energy spectra are not known, in previous studies we have seen general good agreement between simulated and experimental recoil energy distributions for the beam-on dataset~\cite{schueler}, suggesting that the simulated neutron spectrum may be accurate. This simulation suggests typical neutron energies of approximately $0.5$~MeV and very few above $10$~MeV. The source and spectrum of neutrons in the beam-off dataset is unknown. 

\subsection{\label{sec:efficiencies}Efficiency}
At each of the labeled steps of \ptr illustrated in Fig.~\ref{fig:ptr_mc}, some fraction of tracks are removed due to failing quality checks. These checks are:
\begin{description}
\item[a:] successful conversion of hits to energy and position. 
\item[b:] successful completion of SVD prefit ($\theta$ and $\phi$ are finite and defined).
\item[c:] successful slicing and sampling (at least one slice with at least four samples).
\item[e(top):] successful weighted means of $\sigma_L$ and $\sigma_T$ (both are finite and defined).
\item[e(bottom):] successful smeared Bragg fit ($\chi^2_{\text{ndf}}<100$ and $q^{\ptr}<50000$).
\item[f:] successful SVD postfit ($\phi^{\ptr}$ and $\theta^{\ptr}$ are finite and defined).
\end{description}

In both the performance section and in the experimental data studies presented below, we apply all checks listed above and evaluate only the tracks that pass all requirements. The efficiencies for these requirements are summarized in Table~\ref{tab:efficiencies}. Due to preselections in the experimental dataset to discriminate against x-ray recoil candidates, it has fewer low-energy ($E<40$~keV) events. Consequently, the final efficiency for the simulated dataset is somewhat lower.

\begin{table}
\caption{Efficiencies for the quality checks of each step of \ptr for the beam-on sample ($\epsilon_{\text{data}}$) and the simulated sample used in the performance study ($\epsilon_{\text{sim}}$). Step numbers refer to the labeled panels in Fig.~\ref{fig:ptr_mc}. The efficiencies are cumulative, so that the last row gives the total efficiency.}
\label{tab:efficiencies}
\begin{tabular}{lll}
\hline\noalign{\smallskip}
Step & $\epsilon_{\text{sim}}$ & $\epsilon_{\text{data}}$ \\
\noalign{\smallskip}\hline\noalign{\smallskip}
$a$       & 1 & 1 \\
$b$       & 1 & 1 \\
$c$       & 1 & 0.9997 \\
$e(top)$  & 0.9973 & 0.9962 \\
$e(bottom)$  & 0.9683 & 0.9845 \\
$f$       & 0.9473 & 0.9845 \\
\noalign{\smallskip}\hline
\end{tabular}
\end{table}

\subsection{\label{sec:calibration}Calibration}
\begin{figure}
  \includegraphics[width=\columnwidth]{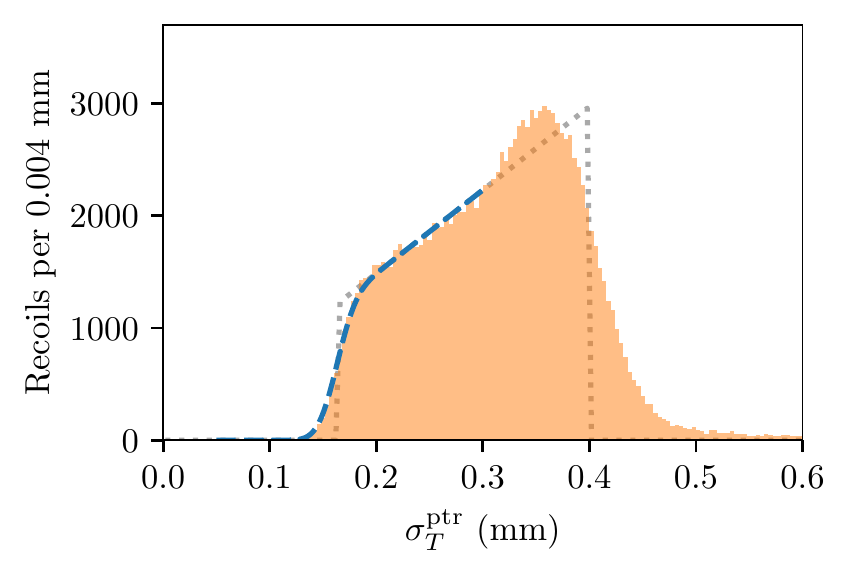}
  \caption{A histogram of the transverse width $\sigma_T^{\ptr}$ (solid orange bars) from the beam-on dataset with $E>100$~keV, with a fit to the leading edge (dashed blue line) using Eq.~\ref{eq:f_sigT} convolved with a Gaussian smearing function. The dotted gray line shows the fitted function over the whole range and without smearing.}
\label{fig:sigmaT_beam_on}
\end{figure}

We begin by demonstrating a novel method of calibrating the values of $A_T$ and $B_T$, using the beam-on dataset. We expect the TPC to be uniformly illuminated by neutrons in $z$ due to the small size of the TPC compared to its distance from the beam pipe (over 1~m). We can then write the event density $f(z)$ of recoils simply as:
\begin{align}
  f(z) =
  \begin{cases}
  \dfrac{N}{z_{\textrm{max}}} & \textrm{if } 0 < z < z_{\textrm{max}},\\[12pt]
  0 & \textrm{otherwise},
  \end{cases}
\end{align}
where $N$ is the total number of events recorded, we have assumed that $z=0$ corresponds to the bottom of the drift volume, and $z_{\textrm{max}}=10.87$~cm is the maximum drift distance. This can be transformed to an event density in $\sigma_T$ using Eq.~\ref{eq:sigmaT_vs_z} and the chain rule:
\begin{align}\label{eq:f_sigT}
  f(\sigma_T) =
  \begin{cases}
  \dfrac{2N\sigma_T}{z_{\textrm{max}}B_T^2} & \textrm{if } A_T < \sigma_T < \sqrt{A_T^2+B_T^2z_{\textrm{max}}}, \\[12pt]
  0 & \textrm{otherwise}.
  \end{cases}
\end{align}
A histogram of $\sigma_T$ values from a uniform source then should have a central, sloping plateau that is linear in $\sigma_T$, points back to the origin, and is surrounded by empty bins. The position of the leading edge of the plateau uniquely determines $A_T$, while the initial height and slope of the plateau uniquely determine $B_T$.  We present in Fig.~\ref{fig:sigmaT_beam_on}, for the first time, a calibration of the diffusion parameters $A_T$ and $B_T$ via a fit to the leading edge of the distribution of $\sigma_T^{\ptr}$, using Eq.~\ref{eq:f_sigT} convolved with a Gaussian smearing function with floating width.  For this calibration we only consider tracks with reconstructed energy over 100~keV due to the presence of a large excess of low-energy recoil candidates near the maximum width. From this fit, we estimate $A_T = \SI{162.35\pm0.19}{\micro\meter}$ and $B_T = \SI{119.17\pm0.24}{\micro\meter}/\si{\centi\meter}$, which are 13\% higher and 11\% lower than the simulated values, respectively. The uncertainties on the parameters come from the fit only, and we find that the fit is insensitive to the choice of fitting endpoint.

We observe a significant residual excess near the high end of the plateau that could be the result of non-primary processes such as activation or (n, $\alpha$)-reactions in the cathode. Nevertheless, qualitative agreement between the model and the rest of the distribution suggests that the underlying assumptions are correct. Other systematic effects may affect the absolute accuracy of $A_T$ and $B_T$, but for fiducialization purposes those errors are not relevant. Consequently, we use this fit to determine the diffusion parameters for fiducialization in subsequent studies.

\subsection{Demonstration: fiducialization}
We use the beam-off dataset to probe the performance of $z_{\text{abs}}$ and fiducialization. We expect two classes of neutron candidates in this dataset: first, \textit{cathode} tracks are nuclei, frequently environmental radon progeny~\cite{radon_drift}, emitted from the cathode that should have a $z_{\text{abs}}$ consistent with the position of the cathode, at $z=10.87$~cm. Second, \textit{uniform background} could come from nuclear recoils generated by externally incident neutrons. This background source might show a flat distribution in $z_{\textit{abs}}$.

\begin{figure}
  \includegraphics[width=\columnwidth]{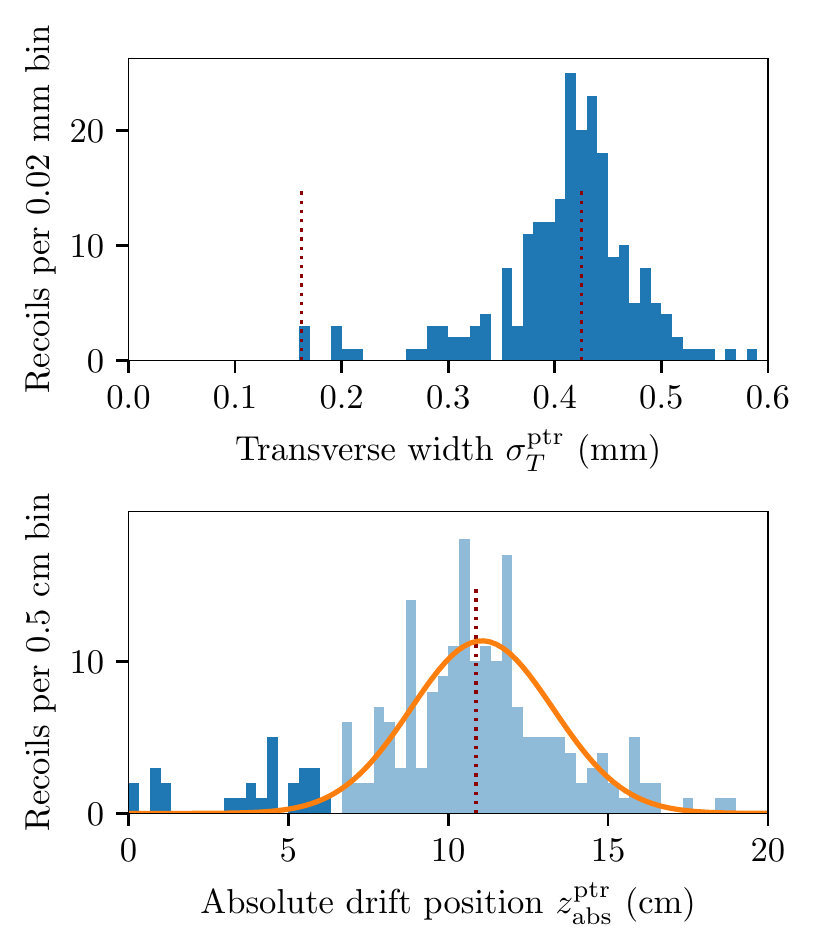}
  \caption{Histograms of the transverse width $\sigma_T$~(top) and derived absolute position \absz~(bottom) for neutron candidates from the beam-off dataset. The expected left and right edge of the $\sigma_T$ plateau from the beam-on calibration of Sec.~\ref{sec:calibration} and the position of the cathode in $z$ are shown with dotted vertical lines. We fit the $z_{\text{abs}}$ distribution with a Gaussian (orange line) to define a $2\sigma$ exclusion region (light blue range). The remaining tracks (dark blue) are considered to be inside the fiducial volume of the TPC.}
\label{fig:absz_baseline}
\end{figure}

In Fig.~\ref{fig:absz_baseline}~(top) we show the distribution of $\sigma_T^{\ptr}$ for the beam-off dataset, for all tracks with $E^{\ptr}>40$~keV. We see a sharp peak near $\sigma_T=\SI{400}{\micro\meter}$, consistent with the right edge of the plateau seen in the beam-on dataset, Fig.~\ref{fig:sigmaT_beam_on}, suggesting that this peak is due to cathode tracks. In addition, we see some evidence of uniform background inside the volume, extending from the expected left edge of the plateau at $A_T=\SI{162.35}{\micro\meter}$. We can conclude that the large majority of the tracks seen during the beam-off run are cathode tracks.

In order to convert the $\sigma_T$ values into \absz, we use the parameters determined in Sec.~\ref{sec:calibration}. We show the resulting distribution of \absz in Fig.~\ref{fig:absz_baseline}. A Gaussian fit to the cathode peak yields a mean position of $z=11.06\pm0.21$~cm, with a width of $2.23\pm0.26$~cm. For comparison, the cathode is at $z=\SI{10.87}{\centi\meter}$. Although in principle the readout plane can be a source of background recoil candidates, we find no conclusive evidence of a source of tracks near $z=0$~cm. We therefore define the volume $z<6.47$~cm to be the fiducial volume and thereby expect to reject $97.7\%$ of cathode tracks (the one-sided Gaussian integral above $-2\sigma$).

While this demonstration is sufficient to validate the basic capabilities of \ptr in determining $z_{\text{abs}}$, we see considerably worse resolution in data ($2.2$~cm) compared to simulation (roughly $1$~cm for $E>40$~keV). This can be attributed to two factors. First, we have applied a calibration from one TPC in the beam-on run to all eight TPCs during the beam-off run several months prior. Better performance could be achieved with independent calibrations for each TPC. Second, cathode tracks are down-going, with a number being very near perfectly vertical. These tracks are difficult to profile and give poor results for $\sigma_T$. 

The small drift distance for the BEAST TPCs also limits the efficiency of the fiducialization. Given a fixed error on $\sigma_T$, we expect the uncertainty of $z_{\text{abs}}$ to grow as the reciprocal of the slope of $\sqrt{z}$. For a 1 meter drift distance more typical of low-background TPCs, we then expect to exclude roughly $2\times\SI{2.2}{\centi\meter}\times\sqrt{\SI{100}{\centi\meter}}/\sqrt{\SI{10}{\centi\meter}}=\SI{14}{\centi\meter}$ due to the fiducialization, a far smaller fractional loss.

\subsection{Demonstration: longitudinal width}\label{sec:demo_sigmaL}

\begin{figure}
  \includegraphics[width=\columnwidth]{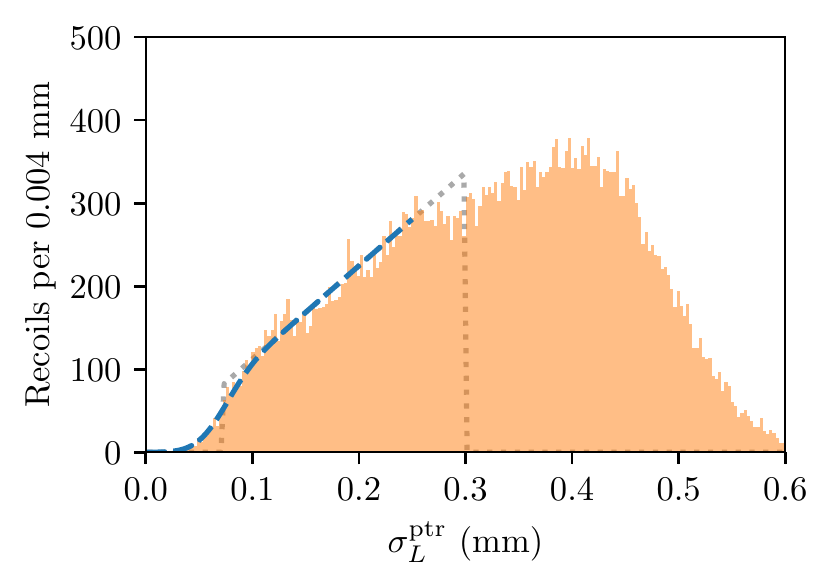}
  \caption{A histogram of $\sigma_L^{\ptr}$ (solid orange bars) from the beam-on dataset with $E>100$~keV and $75^{\circ}<\theta<105^{\circ}$, with a fit to the leading edge (dashed blue line) using Eq.~\ref{eq:f_sigT} convolved with a Gaussian smearing function. The dotted gray line shows the fitted function over the whole range and without smearing.}
\label{fig:sigmaL_beam_on}
\end{figure}

To test \ptr's novel ability to measure the longitudinal width $\sigma_L$ on a track-by-track basis, we look at the distribution of $\sigma_L^{\ptr}$ using the beam-on sample, as in Fig.~\ref{fig:sigmaT_beam_on}, for nearly horizontal tracks (Fig.~\ref{fig:sigmaL_beam_on}). We find the expected sloped plateau, confirming that $\sigma_L^{\ptr}$ depends on $z$ in the expected way. We fit this distribution and find that $A_L=\SI{71.39\pm7.7}{\micro\meter}$ and $B_L=\SI{266\pm13}{\micro\meter}$, which are $27\%$ less than and $107\%$ greater than the values we obtained from simulation, respectively. However, the plateau is far broader than can be explained using the model of Eq.~\ref{eq:f_sigT}.

As a secondary validation, we utilize the fact that $\sigma_L$ and $\sigma_T$ should behave similarly as a function of drift distance. Using Eqs.~\ref{eq:sigmaT_vs_z} and \ref{eq:sigmaL_vs_z}, we can write $\sigma_L^2$ in terms of the diffusion parameters and $\sigma_T^2$:
\begin{align}
  \sigma_L^2 = \left[ A_L^2-\frac{A_T^2B_L^2}{B_T^2}\right] + \frac{B_L^2}{B_T^2}\sigma_T^2.
\end{align}

If the model for $\sigma_L$ is correct, and its determination is robust, then $\sigma_L^2$ will be linearly correlated with $\sigma_T^2$. Using the beam-on dataset with $E>100$~keV and $75^{\circ}<\theta<105^{\circ}$, we find a Pearson correlation coefficient between $\sigma_L^2$ and $\sigma_T^2$ of $0.35$, indicating modest evidence of a linear relationship. 

\begin{figure}
  \includegraphics[width=\columnwidth]{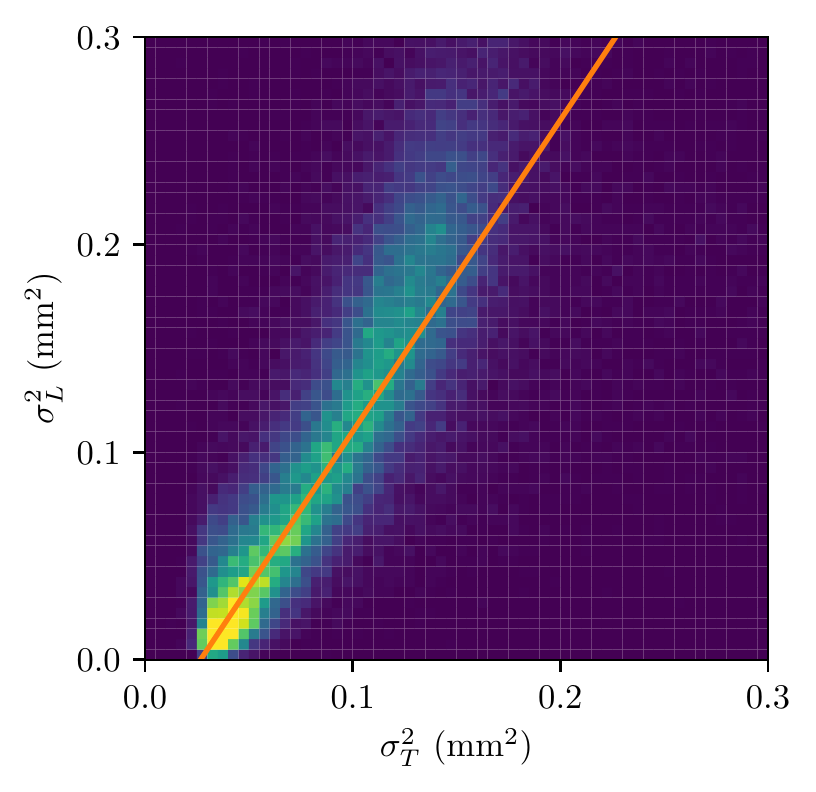}
  \caption{A 2D histogram of $\sigma_L^2$ vs. $\sigma_T^2$ using the beam-on dataset with $E>100$~keV and $75^{\circ}<\theta<105^{\circ}$. The orange line shows the apparent linear relationship in the distribution, and has an offset of $-0.04$~mm$^2$ and a slope of 1.5.}
\label{fig:sigmaL_vs_sigmaT}
\end{figure}

In Fig.~\ref{fig:sigmaL_vs_sigmaT} we show a 2D histogram of $\sigma_L^2$ and $\sigma_T^2$ and find that the main relationship between the two variables is well-defined using a line with offset $\SI{-0.04}{\milli\meter\squared}$ and slope 1.5.  However, the slope should be equal to the ratio of diffusion coefficients $D_L/D_T$, and the observation that this ratio is significantly greater than unity is inconsistent with our expectation that longitudinal diffusion is comparable to transverse diffusion. Additionally, we see a departure from the linear relationship at high $\sigma_T$, presumably related to the excess of events above the inferred $z_{\textrm{max}}$ in Fig.~\ref{fig:sigmaL_beam_on}. These departures from expectation cannot be fully explained by the inclination- and energy-dependent errors in $\sigma_L$, as seen in Fig.~\ref{fig:performance_sigma}, which result in roughly a $10\%$ bias on $\sigma_L^2$ in the selected range of inclinations and energies. 

This poor agreement between expectation and measurement calls into question the assumptions intrinsic to the shell model, most prominently the assumption that the charge pulse develops instantaneously on the pixel. To test this assumption, we perform simulations of induced charge pulse development using the Shockley-Ramo theorem~\cite{ramo}. Following the treatment in Ref.~\cite{pohl}, we approximate the weighting field of the FE-I4b pixel chip with an analytic function that takes into account the collection gap as well as the size and relative placement of the electrodes. We consider the electrodes to have the dimensions of the metallization layer deposited on each pixel ($100\times$\SI{10}{\micro\meter\squared}), and to be spaced according to the pixel pitch. 

We observe two phenomena in this simulation. First, a \textit{direct pulse} is induced on an electrode by electrons traveling between the GEM and the electrode. Second, an \textit{indirect pulse} is induced by electrons traveling between the GEM and neighboring electrodes. While the direct pulse results in net induced charge, the indirect pulse integrates to zero by the end of drift. However, it is not zero during drift.

\begin{figure}
  \includegraphics[width=\columnwidth]{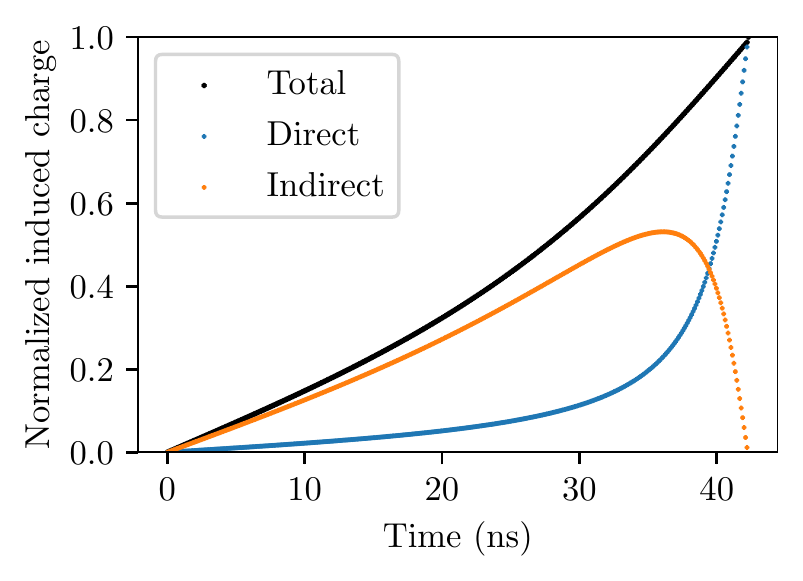}
  \caption{A simulation of induced charge development on a pixel of the BEAST TPCs, using the Shockley-Ramo theorem and assuming a uniform charge distribution with no diffusion. Direct charge is induced by charges drifting to the pixel, while indirect charge is induced by charge drifting to neighboring pixels. The charge drifts over the \SI{2.2}{\milli\meter} collection gap and induces charge on a \SI{100}{\micro\meter} electrode.}
\label{fig:ramo}
\end{figure}

We find that the weighting field is concentrated near the electrodes, and therefore the direct pulse develops on a timescale of a few nanoseconds. This is small compared to the threshold-crossing time resolution of the FE-I4b. However, the indirect pulses are not small, and they develop earlier than direct pulses. In fact, With BEAST TPC geometry we find that threshold crossing typically occurs long before the direct pulse develops significantly (see Fig.~\ref{fig:ramo}), and it is possible for the indirect pulse to advance the crossing time by up to two clock cycles for a slowly developing charge cloud.

Because high-charge pixels are typically surrounded by other high-charge pixels, the advancement of the threshold-crossing time due to the indirect pulse depends on the local charge density. This mechanism will lead to a deeper charge shell than expected. When fit with the shell model, the extracted value for the longitudinal width $\sigma_L$ will be biased high, particularly for tracks with large transverse width $\sigma_T$. This is consistent with the effects we see on experimental data, and, because we have assumed instantaneous charge collection in simulation, explains why the longitudinal width $\sigma_L$ estimate is unbiased in simulation but not in experimental data. However, we expect the $z$ coordinate of the track center for a given slice, $z_0$, to be unbiased.

We conclude that \ptr is likely sensitive to $\sigma_L$ on a track-by-track basis, but the estimation is not unbiased. A multitude of factors contribute to the impact of the indirect pulse on the charge development on a pixel, and we do not attempt to parameterize and model all of these effects here. Rather, we note that the bias is due to Shockley-Ramo considerations, therefore it is entirely a function of detector geometry. In particular, closing the collection gap to dimensions comparable to the pixel pitch virtually eliminates the indirect pulse. We discuss implications for the future use of \ptr in Sec.~\ref{sec:extending_dispersion}.

\subsection{Demonstration: energy and length resolution}\label{sec:E_and_L}
\begin{figure}
  \includegraphics[width=\columnwidth]{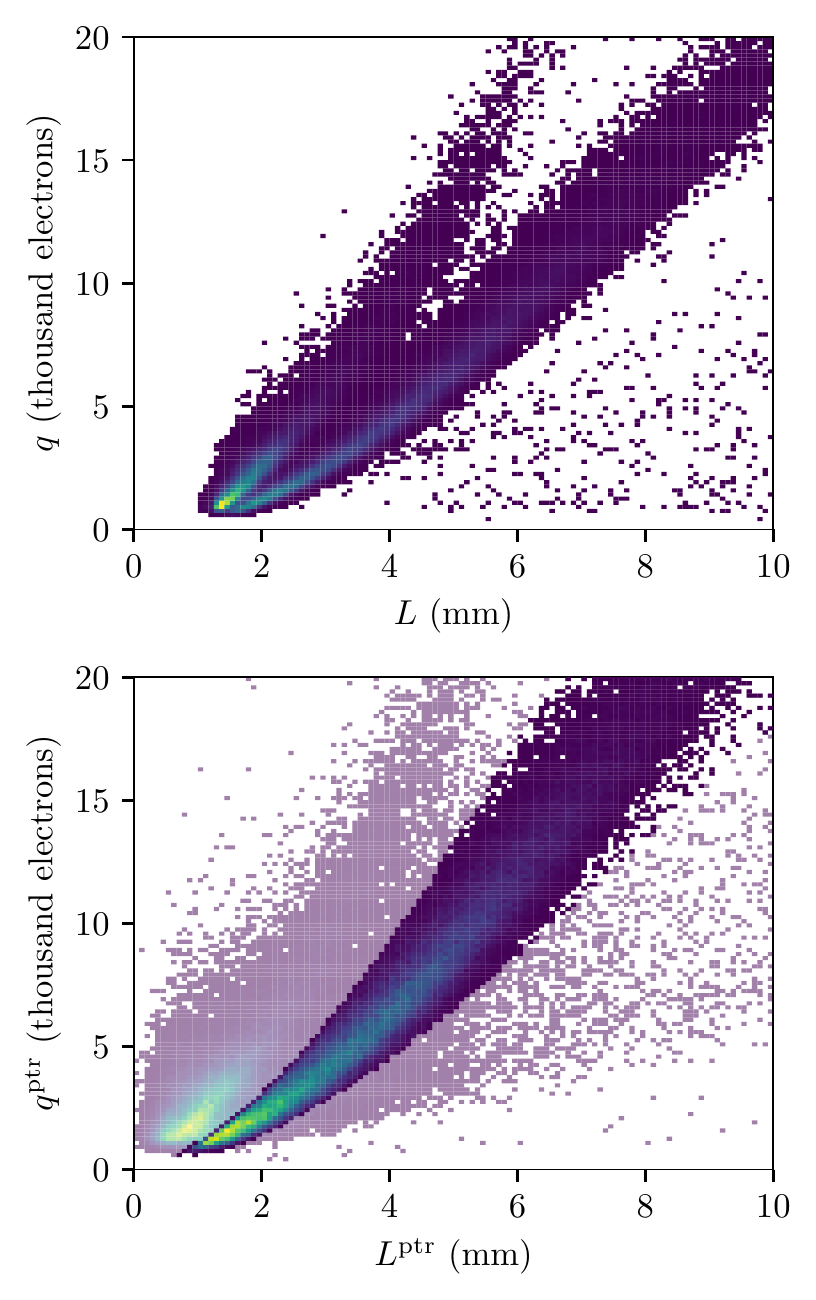}
  \caption{2D histograms of $L$ vs. $q$ for the benchmark method (top) and for \ptr (bottom), using the beam-on sample. In both plots, the bottom band corresponds to alpha nucleus recoils, while the top band corresponds to C/O nucleus recoils. For \ptr, the partially transparent events fall outside of a selection on the floating smeared Bragg scale factor $0.6 < s < 1.2$.}
\label{fig:L_vs_q}
\end{figure}

We investigate the performance of \ptr in removing charge and length biases by plotting reconstructed charge versus length for the beam-on data (Fig.~\ref{fig:L_vs_q}). Compared to the benchmark method, we observe a significant shift to lower lengths for alpha recoils (the bottom band in the plots), consistent with \ptr deconvolving diffusion. We also see that high-energy events are shifted to higher energies, suggesting that \ptr has also successfully recovered charge lost to saturation. 

In the smeared Bragg fit we have introduced a floating scale factor $s$ that has a value of exactly 1 if the recoil matches the expected $\textrm{d}E(x)/\textrm{d}x$ for alpha recoils. This factor accommodates both fluctuations in the effective gain, and, because we have not implemented Bragg models for non-alpha recoils, it also permits fits to recoils of other species. Without this free parameter, the energy-length relationship would be constrained to that of the alpha recoil parameterization, and no discrimination between species would be possible except by $\chi^2$ value. In Fig.~\ref{fig:L_vs_q}(bottom) we show the results with and without a selection on this scale factor $0.6 < s < 1.2$. This selection is asymmetric about 1.0 due, presumably, to a somewhat lower effective gain compared to the calibrated value. We see that such a selection provides a simple means to isolate recoils of a specific species, validate the gain calibration, and filter out events with low-quality fits. 
  
As an estimate of the $\textrm{d}E/\textrm{d}x$ resolution of \ptr, we evaluate the standard deviation of $q^{\ptr}$ in narrow bins of $L^{\ptr}$. We find that the resolution of \ptr is about $15\%$ worse than the benchmark method across all lengths, while it is around $6\%$ worse in simulation. The degraded $\textrm{d}E/\textrm{d}x$ resolution with \ptr may be attributed to range straggling; the track model enforces the statistical mean Bragg behavior, and thus differences in track length between tracks of the same energy are not accommodated. Thus \ptr may be considered over-constrained, a feature that is beneficial for removing biases and improving the head-tail efficiency, but not for high-resolution charge and length determinations. The errors caused by imposing the statistical mean Bragg behavior are particularly sensitive to the extreme end of the recoil. Small errors in the position of the last few ionizations by SRIM could explain the relatively large difference between the energy resolutions on simulated and experimental data. 

In order to improve the charge and length resolution of \ptr, it may be possible to include a free parameter in the Bragg parameterization that accommodates changes in the endpoint without distorting the peak. This will decouple charge and length and may improve resolution. Alternatively, the sum of the slice charges may be a minimally biased and better-resolution charge estimator, which will be best demonstrated using data from a monochromatic neutron source.

An important application of charge and length is the identification of recoil species, often accomplished by defining regions in the $L$ vs. $E$ plane. This does not require unbiased estimates for charge and length. Therefore, it may be optimal to use existing methods for particle identification purposes, then use \ptr to obtain an estimate of the unbiased energy or length of each remaining track. A comprehensive evaluation and optimization of the performance of \ptr for particle identification purposes is a subject for future study.

\subsection{Demonstration: head-tail}
For the beam-on dataset, fast neutrons are produced when beam particles or collision products collide with the walls of the SuperKEKB beam pipe, ultimately stimulating the emission of neutrons via the giant dipole resonance~\cite{phase1}. With the TPCs placed about 1~m from and oriented nearly parallel to the beam pipe, the neutrons should constitute a point source in $\phi$. The recoil distribution should be broad and centered on the position of the beam pipe ($\phi=\pm180$), with a non-peaking background due to neutron production outside the beam pipe or neutrons scattered by surrounding material.

\begin{figure}
  \includegraphics[width=\columnwidth]{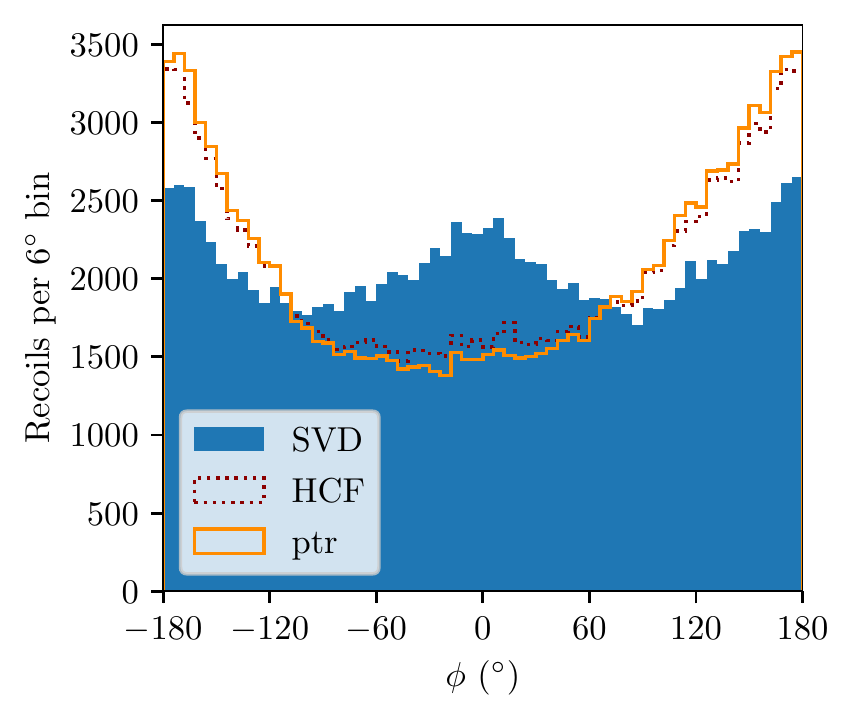}
  \caption{Histograms of $\phi$ from the SVD prefit (solid blue) and after flipping according to HCF (dotted line) and \ptr (solid line) using the beam-on dataset and alpha recoil candidates identified by the selection shown in Fig.~\ref{fig:L_vs_q}(bottom). The beam pipe is located at $\phi=\pm180^{\circ}$, and the peak in the prefit distribution at $\phi=0^{\circ}$ is presumably due to head-tail swaps. With \ptr, we see a complete reassignment of these swapped tracks to the correct peak.}
  \label{fig:hotspot}
\end{figure}

Due to the head-tail ambiguity of the SVD prefit, we expect the point source in $\phi$ to be split evenly between $\phi$ and the remainder of $(\phi+180^{\circ})/360^{\circ}$. This can be seen in Fig.~\ref{fig:hotspot}, where the beam pipe is located at $\pm180^{\circ}$; the peak at $\phi=0^{\circ}$ is presumably due to head-tail swaps of tracks from neutrons originating in the beam pipe. Head-tail sensitivity is needed to confirm this hypothesis. 

We use \ptr to determine the correct head-tail assignment for alpha recoil candidates ($0.6 < s < 1.2$), shown in Fig.~\ref{fig:hotspot}. We find that our hypothesis is correct: we observe a significant peak at $\pm180^{\circ}$, corresponding to the position of the beam pipe, with no remaining excess events at $\phi=0^{\circ}$, corresponding to the position of a concrete wall. With HCF, we find a small remnant of excess events at $\phi=0^{\circ}$, suggesting that it is not as efficient at determining head-tail in this sample. 

To probe the head-tail performance, we fit this histogram with a function consisting of a constant plus two Gaussians, one centered at 0 and the other centered at $\pm180^{\circ}$, with a shared width. We take the head-tail efficiency to be the ratio of fitted yields $N_{180}/(N_{180}+N_{0})$. We find no evidence of mis-assigned head-tail using \ptr, with $\epsilon_{\text{ht}}^{\ptr}=1.0000^{+0}_{-0.0001}$ (fit uncertainty only). For HCF, we find $\epsilon_{\text{ht}}^{\text{HCF}}=0.979\pm0.031$, also consistent with 1.

These efficiencies are far above both what we found in the MC performance studies, and in a recently submitted paper~\cite{hedges} that used the HCF method. This is most likely due to a poor understanding of the angular distribution of the neutrons that do not come directly from the beam pipe. Consequently, we can make no quantitative claims about the head-tail efficiency of \ptr. However, we consider this test to constitute strong qualitative evidence that \ptr is both highly efficient at assigning the correct head-tail, and that it outperforms the HCF method. Future experimental tests using a point source in a clean and well-controlled environment will be required to evaluate the head-tail efficiency more rigorously.

\section{\label{sec:conclusions}Discussion and implications}
  We have seen that the two key technical limitations for low-energy recoil reconstruction performance in the BEAST TPCs are charge integration and resolution effects due to diffusion, amplification, and digitization. We summarize these here and comment on the implications for other detector technologies.

\subsection{\label{sec:extending_pileup}Integration effects}
Charge integration is largely necessitated by the high drift velocity of electrons. We see significant adverse impacts on track reconstruction performance in the following ways:
\begin{itemize}
\item Loss of charge through pixel saturation, especially for highly inclined tracks.
\item Loss of length and angle sensitivity for highly inclined tracks.
\item Loss of charge structure information transverse to the tracks in $z$ ($\sigma_L$) for all tracks.
\item Loss of charge structure information along the track in $z$ (Bragg) for inclined tracks. 
\item Biasing of head charge fraction for highly inclined tracks.
\item Diminishing of ability to fiducialize via $\sigma_T$ for highly inclined tracks. 
\end{itemize}
We find that \ptr can significantly mitigate all of these impacts. However, the optimal performance will be achieved in detectors where the integration time is small compared to the typical time of arrival difference between successive charges. This can be achieved with NID TPCs, which can measure the time of arrival of each charge in the primary track individually, thereby obviating the need to model and deconvolve integration dynamics. 

\subsection{\label{sec:extending_dispersion}Resolution}
Although we have shown that \ptr can successfully deconvolve diffusion and other resolution effects, the low-energy limit is still determined by the prefit accuracy, which is ultimately limited by the amplification and readout resolution. Therefore we expect to be able to extend the low-energy limit significantly by using a detector with the amplification integrated into a very high-resolution readout. Such \textit{InGrid} detectors~\cite{ingrid} have been developed recently, for example GridPix~\cite{gridpix}, in which a Micromegas mesh is directly mounted onto a TimePix3 pixel chip. With this readout, each primary charge is mapped to a single pixel. The amplification and readout resolution term therefore is in principle determined solely by the pixel pitch ($A_T\approx50/\sqrt{12}~\si{\micro\meter}$)~\cite{gridpix2}. Additionally, due to the very short collection gap, these sensors should be immune to threshold-crossing-time biases due to indirect charge pulses (Sec.~\ref{sec:demo_sigmaL}).

With minimal amplification and readout resolution effects, and especially for TPCs with large drift length, drift diffusion becomes the limiting effect. This can be substantially reduced by using NID, but for low-rate applications such as directional dark matter searches this will typically be negated by an expansion of the drift length. Therefore, deconvolution of diffusion will still be necessary to achieve optimal performance. A detector coupling NID with GridPix has recently been demonstrated~\cite{nikhef}. Using this technology, the low-energy threshold for directionality may ultimately be determined by the transverse straggling of recoils. In this regime, the effective charge weighting of \ptr's refit and its ability to reconstruct transverse straggling may significantly improve vector tracking compared to simpler track reconstruction methods.

\subsection{\label{sec:other_technologies}Other technologies}
Other detectors with alternative amplification and readout technologies will similarly be limited by diffusion and digitization effects. We expect that \ptr can be easily adapted for use in a wide range of detectors. A necessary modification for detectors without charge integration would be to replace the charge shell fit with a profile fit in $z$. Other effects, such as charge spreading, must be built into the model to be properly deconvolved. 

\section{\label{sec:extending_optimal}Conclusions}
We have described a model for track development and digitization in HD TPCs and used this model to build an algorithm that effectively deconvolves diffusion and digitization effects from detected nuclear recoil tracks below a transverse aspect ratio of \lst$=2$. We have shown that this algorithm is capable of recovering information lost through charge integration such that the detected tracks can now be considered to be 3D up to high track inclinations. This leads to a major improvement in the efficiency of the head-tail determination relative to existing methods (to $0.78$ for a track aspect ratio of \lst$=1.5$ on the readout plane), and newly unbiased estimates of energy, length, transverse diffusion width, and the absolute $z$ position. Additionally, we have provided preliminary evidence that it is possible to exploit the threshold-crossing time structure of a track to extract the longitudinal diffusion width. For an arbitrary detector, we find that \ptr should decrease the low-energy threshold for accurate track reconstruction relative to existing techniques.  

The benefits of \ptr are counterbalanced by some limitations compared to existing techniques. These include inferior charge resolution and inadequate simulation and modeling of pulse development effects. Both of these effects can be corrected within the \ptr framework with further study. In addition, further dedicated experimental tests are required to fully validate the real-world performance of \ptr.

\begin{acknowledgements}
We thank Cosmin Deaconu for help with recoil simulation. We thank Tommy Lam and Annam L\^e for technical assistance in construction and testing. We thank David-Leon Pohl and Jens Janssen for special contributions to the readout firmware  and David-Leon Pohl again for consultations regarding Shockley-Ramo considerations. We acknowledge support from the U.S. Department of Energy (DOE) via Award Numbers DE-SC0007852, DE-SC0010504.
\end{acknowledgements}

\printbibliography

\end{document}